\documentclass[aip,pof,preprint,amsmath,amssymb]{revtex4-1}
\usepackage{graphicx}% Include figure files
\usepackage{dcolumn}% Align table columns on decimal point
\usepackage{bm}% bold math
\begin{document}

\title[Zero-{$\Pr$} rotating convection]{Zero-Prandtl-number convection with slow rotation}% Force line breaks with \\

\author{Priyanka Maity}
 \affiliation{Department of Physics, Indian Institute of Technology, Kharagpur-721~302, India}
\author{Krishna Kumar}
 \email{kumar@phy.iitkgp.ernet.in}
  \affiliation{Department of Physics, Indian Institute of Technology, Kharagpur-721~302, India}

\date{\today}% It is always \today, today,

\begin{abstract}
We present the results of our investigations of the primary instability and the flow patterns near onset in zero-Prandtl-number Rayleigh-B\'{e}nard convection with uniform rotation  about a vertical axis.  The investigations are carried out using direct numerical simulations of the  hydrodynamic equations with stress-free horizontal boundaries in rectangular boxes of size $(2\pi/k_x) \times (2\pi/k_y) \times 1$ for different values of the ratio $\eta = k_x/k_y$. The primary instability is found to depend on $\eta$ and $Ta$. Wavy rolls are observed at the primary instability for smaller values of  $\eta$ ($1/\sqrt{3} \le \eta \le 2$ except at $\eta = 1$) and for smaller values of $Ta$. We observed K\"{u}ppers-Lortz (KL) type patterns at the primary instability for $\eta = 1/\sqrt{3}$ and $ Ta \ge 40$. The fluid patterns are found to exhibit the phenomenon of bursting, as observed in experiments [Bajaj et al. Phys. Rev. E {\bf 65}, 056309 (2002)]. Periodic wavy rolls are observed at onset for smaller values of $Ta$, while KL-type patterns are observed for $ Ta \ge 100$ for $\eta =\sqrt{3}$. In case of $\eta = 2$, wavy rolls are observed for smaller values of $Ta$ and KL-type patterns are observed for $25 \le Ta \le 575$. Quasi-periodically varying patterns are observed in the oscillatory regime ($Ta > 575$). The behavior is quite different at $\eta = 1$. A time dependent competition between two sets of mutually perpendicular rolls is observed at onset for all values of $Ta$ in this case. Fluid patterns are found to burst periodically as well as chaotically in time.  It involved a homoclinic bifurcation.  We have also made a couple of  low-dimensional models to investigate bifurcations for $\eta = 1$, which  is used to investigate the sequence of bifurcations. 
\end{abstract}

\keywords{Rotating zero-Prandtl-number convection, pattern-forming instability, self-tuned waves, homoclinic bifurcation}
\maketitle
\section{Introduction}
Thermal convection in low-Prandtl-number ($Pr \ll 1$) fluids~\cite{busse_1972, proctor_1977, clever_busse_1981, shew_lathrop_2005, chiffaudel_etal_1987, fbb_1987, croquette_1989, libchaber_etal_1982, liu_ahlers_1996, xi_etal_1997, egolf_etal_2000, thual_1992, pal_etal_2009, mishra_etal_2010, pal_etal_2013} is widely investigated due to its relevance in the study of convective zones in stars and the Sun ($Pr \sim 10^{-8}$), Earth's molten core~\cite{shew_lathrop_2005} ($Pr \sim 10^{-1}$), liquid metals~\cite{chiffaudel_etal_1987} ($Pr$ $\sim$ $10^{-3}$ - $10^{-2}$), and pattern-forming instabilities~\cite{fbb_1987, croquette_1989, libchaber_etal_1982, liu_ahlers_1996, xi_etal_1997, egolf_etal_2000, thual_1992, pal_etal_2009, mishra_etal_2010, pal_etal_2013}.  Rayleigh-B\'{e}nard convection~\cite{chandrasekhar_book}, which consists of a thin horizontal fluid layer uniformly heated from below and uniformly cooled from above, is a simplified version to study the complexities of thermal convection. The convection in a Boussinesq fluid is governed by two nonlinearities: the self-interaction of the velocity field $(\boldsymbol{v}\cdot\boldsymbol{\nabla})\boldsymbol{v}$ and the advection of temperature by the velocity field $({\boldsymbol{v}\cdot\boldsymbol{\nabla}})\theta$. Proctor~\cite{proctor_1977} suggested two different flow regimes for the convection near the instability onset in low-Prandtl-number fluids: the inertial and the viscous regimes. If $(\boldsymbol{v}\cdot\boldsymbol{\nabla})\boldsymbol{v}$ is nonzero but compensated by the pressure gradient, the flow dynamics is described by the nonlinear term $({\boldsymbol{v}\cdot\boldsymbol{\nabla})}\theta$. The convective heat flux ($Nu -1$) across the fluid layer could be large and independent of $Pr$. The corresponding convection is said to be in the inertial regime. If the convective heat flux is small and $Nu -1 \sim {Pr}^2$, the corresponding convection is said to be in the viscous regime. Chiffaudel, Fauve \& Perrin~\cite{chiffaudel_etal_1987} established the existence of these two flow regimes experimentally. The asymptotic form of the Boussinesq equations for Rayleigh-B\'{e}nard convection in the zero-Prandtl-number ($Pr \rightarrow 0$) limit was proposed by Spiegel~\cite{spiegel_1962} to simplify the study of convective turbulence in an astrophysical context. The nonlinear term  $({\boldsymbol{v}\cdot\boldsymbol{\nabla})}\theta$ is neglected in this limit, and the thermal convection is described by the term $(\boldsymbol{v}\cdot\boldsymbol{\nabla})\boldsymbol{v}$. This limit is therefore the viscous regime of convection. The only nonlinearity $(\boldsymbol{v}\cdot\boldsymbol{\nabla})\boldsymbol{v}$ is unable to saturate two-dimensional (2D) rolls, as the 2D growing rolls become the exact nonlinear solution with {\it stress-free} horizontal boundaries. This makes the $Pr \rightarrow 0$ limit with stress-free boundary conditions a singular limit. Thual~\cite{thual_1992} showed the saturation of thermal convection in direct numerical simulations (DNS) of three-dimensional (3D) flows in this limit with stress-free boundaries. A simple model for a possible saturation mechanism of the instability was suggested by Kumar, Fauve \& Thual~\cite{kft_1996}, which also showed the possibility of critical bursting~\cite{kumar_etal_2006} at the primary instability. Recent studies~\cite{pal_etal_2009} show several interesting instabilities including a homoclinic bifurcation close to the primary instability. The strong similarity of convective flows~\cite{mishra_etal_2010, pal_etal_2013} for very small values of $Pr$ with those of Spiegel's equations~\cite{thual_1992, pal_etal_2009} confirm the existence of a smooth $Pr \rightarrow 0$ limit.

Rotation about a vertical axis~\cite{chandrasekhar_book, veronis_1968, kl_1969, kuppers_1970, rossby_1969, riahi_1992, goldstein_etal_1994, cb_2000, cm_2000, dawes_2001a, dawes_2001b, sanchez_etal_2005, lopez_marques_2009, pharasi_kumar_2013} introduces the centrifugal force and the Coriolis force. Both the forces act along the horizontal plane. The centrifugal force~\cite{lopez_marques_2009, becker_etal_2006} introduces large scale circulation into the flow and modifies the temperature profile of the basic state.  The Froude number $Fr$, which is the ratio of the centrifugal force and the buoyancy force, becomes an important parameter for the fluid flow. Lopez and Marques~\cite{lopez_marques_2009} investigated the effects of centrifugal force on fluid flow in detail.  The Coriolis force linearly couples the vertical velocity and the vertical vorticity. This delays the onset of Rayleigh-B\'{e}nard convection~\cite{chandrasekhar_book}, affects the heat flux~\cite{rossby_1969, riahi_1992, julien_etal_jfm_1996, julien_etal_pre_1996, liu_ecke_2009, king_etal_2009, stevens_etal_prl_2009, pharasi_etal_2011, king_etal_2013} across the fluid layer, and breaks the mirror-symmetry of the convective flow even at small rotation rates. The Coriolis force also affects the turbulent spectra~\cite{pharasi_etal_2014}. Chandrasekhar~\cite{chandrasekhar_book} analyzed the linear problem considering only the Coriolis force and showed that the infinitesimal convection always appeared as stationary convection, if $Pr > 0.677$. He predicted oscillatory convection at onset if $Pr < 0.677$ and if $Ta$ is greater than a critical value $Ta_{\circ} (Pr)$. K\"{u}ppers \& Lortz (KL)~\cite{kl_1969, kuppers_1970} investigated the nonlinear convection and showed that the flow was always unsteady at onset if $Ta$ is greater than a threshold $Ta_{KL} (Pr) < Ta_{\circ} (Pr)$. A set of rolls is then replaced by a new set of rolls of the same wavelength but oriented approximately at an angle of $60^{\circ}$ with the old set~\cite{busse_heikes_1980, niemela_donnelly_1986} at the onset. Recent studies on pattern-forming instabilities~\cite{cb_2000, lopez_marques_2009, bajaj_etal_1998, bajaj_etal_2002, sanchez_etal_2005, sprague_2006, scheel_etal_2010} showed interesting patterns. Square patterns~\cite{bajaj_etal_1998, sanchez_etal_2005, scheel_etal_2010} were also observed in presence of rotation. Bajaj etal.~\cite{bajaj_etal_2002} observed bursts of finite-amplitude rolls which appeared and then disappeared at almost regular intervals. Clever \& Busse~\cite{cb_2000} also observed bursting of the convective heat flux in simulations of oscillatory convection with no-slip boundary conditions, while Dawes~\cite{dawes_2001a} found bursting in a model of convection near a bicritical point with stress-free boundary conditions. Not much is known, however, about the phenomenon of bursting of patterns in the regime of stationary convection in a rotating Rayleigh-B\'{e}nard system.

We present in this article the results of a detailed investigation of the fluid patterns near the primary instability in a Rayleigh-B\'{e}nard system rotating uniformly but slowly about a vertical axis in the $Pr \rightarrow 0$ limit. This work is motivated by our curiosity concerning the effect of Coriolis force on the pattern dynamics close to the onset of Rayleigh-B\'{e}nard convection in very low-Prandtl-number fluids. The purposes of this paper are twofold. The first is to investigate K\"{u}ppers-Lortz (KL) instability in very low-Prandtl-number fluids. The second purpose is to explore the effects of the Coriolis force on the phenomenon of critical bursting, which is observed in low-Prandtl-number fluids. We carried out direct numerical simulations (DNS) of the hydrodynamical system with {\it stress-free} boundary conditions in a three-dimensional simulation box of size $2\pi/k_x \times 2\pi/k_y \times 1$ for a wide range of Taylor numbers $Ta$. The ratio $\eta = k_x/k_y$ was varied in a range $1/\sqrt{3} \le \eta \le 10$. The convection is found to be always three-dimensional and time dependent at the primary instability. Irregular bursts of waves were excited along the roll axis in larger boxes, whenever the amplitude of rolls exceeded a critical value for $\eta > 2$. This led to the possibility of chaotic bursting of oblique rolls for $\eta > 2$. For $\eta = 2$, we observed wavy rolls for $Ta \le 10$ and chaotic oblique rolls for $Ta > 10$ near the instability onset. We observed an interesting competition between two sets of  mutually perpendicular rolls in a square simulation box ($\eta = 1$), which shows irregular and regular bursting of fluid patterns. A sequence of inverse homoclinic gluing, inverse Hopf and inverse pitchfork bifurcations was also observed in a small square box, when the Rayleigh number $Ra$ was raised in small steps for a fixed value of $Ta$. Fluid patterns showed periodic as well as chaotic bursting. K\"{u}ppers and Lortz (KL) patterns~\cite{kl_1969} are observed at the onset for certain range of $Ta$, which depend on the ratio $\eta$. For $\eta \ge 4$, the convection is found to be chaotic at the onset even for smaller values of $Ta$. A set of wavy rolls is replaced irregularly in time by a new set of oblique wavy rolls of different wavelength at onset. We have also constructed a couple of low-dimensional models for $\eta = 1$, which captures qualitatively the flow patterns observed in DNS for smaller values of $Ta$ and the reduced Rayleigh number $r$.

\section{Hydrodynamic problem}
We consider a thin horizontal layer of Boussinesq fluid of thickness $d$, kinematic viscosity $\nu$, thermal expansion coefficient $\alpha$, thermal diffusivity $\kappa$, rotating uniformly with angular velocity ${\boldmath{ \Omega}}$ about a vertical axis. The fluid layer is heated uniformly from below and cooled uniformly from top to maintain an adverse temperature gradient $\beta$ across the fluid layer. The Froude number $Fr = \Omega^2 L/g$ is always less than $2.2 \times 10^{-3}$ for the rotation rates considered here ($Ta \le 700$). The effects of centrifugal force are therefore ignored here.
We have called the rotation rates slow, as we have considered the simulations for smaller values of $Ta$ such that the effects of the centrifugal force may be ignored. It is then possible to have stationary conduction as a basic state in the rotating frame of reference. The convective flow, in the Boussinesq approximation, is then governed by the following set of dimensionless equations: 
\begin{equation}
\partial_{t}\boldsymbol{v}+(\boldsymbol{v\cdot\nabla})\boldsymbol{v} = 
    -\boldsymbol{\nabla}p + Ra\theta{\boldsymbol{\lambda}} + \nabla^{2}\boldsymbol{v} 
 + \sqrt{Ta}({\boldsymbol{v\times}{\boldsymbol{\lambda}}}),\label{ns}
\end{equation}
\begin{equation} 
 Pr[\partial_{t}\theta + (\boldsymbol{v\cdot\nabla})\theta] = \nabla^{2}\theta + v_3 \label{temp},
\end{equation}
\begin{equation} 
 \boldsymbol{\nabla \cdot v} = 0.\label{cont}
\end{equation}
\noindent 
The convective flow is then governed by three dimensionless numbers: (1) Rayleigh number $Ra = g \alpha \beta d^4/(\nu \kappa)$, which is a measure of the buoyancy force, (2) Taylor number $Ta = 4 \Omega^2 d^4/\nu^2$, which is proportional to the square of rotation rate, and (3) Prandtl number $Pr = \nu/\kappa$, which is the ratio of the thermal and viscous diffusive time scales.  
In absence of rotation ($Ta = 0$), the hydrodynamic system (equations~\ref{ns}-\ref{cont}) reduces to Spiegel's equations~\cite{spiegel_1962} in the limit of $Pr \rightarrow 0$. The temperature field is then slaved to the vertical velocity in this limit. We consider thermally conducting and stress-free bounding surfaces. The stress-free conditions are relevant for boundaries between two fluids of large viscosity difference. They were realized in experiments by Goldstein and Graham~\cite{goldstein_graham_1969}. The stress-free boundary conditions are also appropriate for analytical work (e.g. amplitude equations) and constructing accurate low-dimensional models, since trigonometric eigenfunctions simplify algebra.  We then have $\theta$ $=$  $\partial_{z}v_{1}$ $=$ $\partial_{z}v_{2}$ $=$ $v_{3} = 0$ on the horizontal surfaces, which are located at $z=0$ and $1$. All the convective fields are assumed to be periodic on the horizontal plane. 

\section{Linear stability analysis}
The linear stability results in the limit of vanishing Prandtl number may be obtained, following Chandrasekhar~\cite{chandrasekhar_book}, in terms of the vertical velocity $v_3$. The relevant equation in the $Pr \rightarrow 0$ limit reads as:
\begin{equation}
\left[ \nabla^2 \left\{ \left(\partial_t - \nabla^2 \right)^2 \nabla^2 + Ta D^2 \right\}  + Ra \left( \partial_t - \nabla^2 \right)\nabla_H^2  \right] v_3 (x,y,z,t) = 0, \label{ls_eq}
\end{equation}
where $D \equiv \partial_z$. The vertical velocity is expanded in normal modes, given by,
\begin{equation}
v_3 (x, y, z, t)= W(z) \exp{\left[ i(k_{x} x + k_{y} y) + \sigma t\right]}, \label{normal_modes}
\end{equation}
where $k = \sqrt{k_x^2 + k_y^2}$ is the horizontal wavenumber. Inserting equation~\ref{normal_modes} in equation~\ref{ls_eq}, and using the trial solution $W(z) = A \sin{(\pi z)}$, which is compatible with stress-free boundary conditions, we arrive at the following stability condition:
\begin{equation}
\left[ \left( \pi^2 + k^2 \right) \left\{ \left(\sigma + {\pi}^2 + k^2 \right)^2 \left({\pi}^2 + k^2\right) + {\pi^2}Ta \right\} - {k^2} Ra \left(\sigma + {\pi}^2 + k^2 \right) \right]= 0. \label{sigma_eqn}
\end{equation}
The roots of the quadratic equation~(\ref{sigma_eqn}) for $\sigma$ are the linear growth rates $\sigma_{\pm}$ of the perturbations. They are given by,
\begin{equation}
\sigma_{\pm} = \frac{-2(\pi^2 + k^2)^3 + k^2{Ra}\pm \sqrt{{(k^2 Ra)^2 - 4 {\pi}^2 (\pi^2 + k^2)^3 Ta}}}{2(\pi^2 + k^2)^2} \label{growth_rate}
\end{equation}
\noindent The two growth rates $\sigma_{+}$ and $\sigma_{-}$ may either be real, or may form a complex conjugate pair. 

\begin{figure}[h]
 \begin{center}
\includegraphics[height=8 cm, width=14 cm]{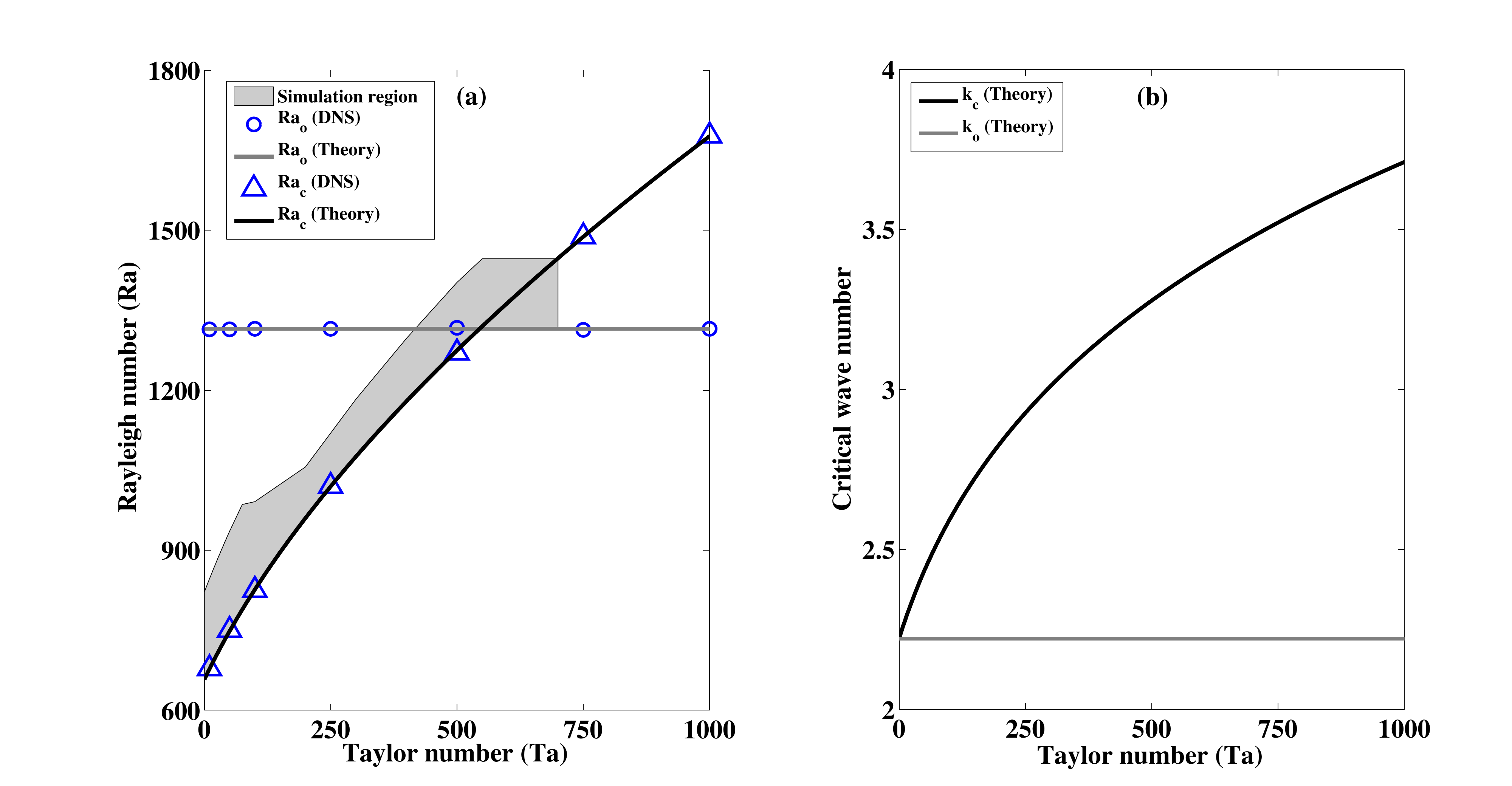}
\caption[short]{Linear stability analysis in the $Pr \rightarrow 0$ limit. (a) The variation of critical Rayleigh numbers $Ra_c (Ta)$ (black curve) for the stationary convection and $Ra_{\circ} (Ta)$ (gray curve) for the oscillatory convection with Taylor number $Ta$. (b) The variation of critical wave numbers $k_c (Ta)$ (black curve) for the stationary convection and $k_{\circ} (Ta)$ (gray curve) for the oscillatory convection with $Ta$. $Ra_{\circ}$ and $k_{\circ}$ are independent of $Ta$ in this limit. The symbols $\triangle$  and $\circ$ are data obtained from DNS of full hydrodynamic system for $Ra_c (Ta)$ and $Ra_{\circ}$, respectively. The shaded area shows the region of the $Ra - Ta$ plane where simulations are done.}
\label{lin_stability}
 \end{center}
\end{figure}

\subsubsection{Stationary convection}
Both the growth rates  are real, if $(k^2 Ra)^2 \geq 4 \pi^2 (\pi^2+k^2)^3 Ta$. 
Setting the larger value $\sigma_{+} = 0$ yields the condition for stationary convection. The critical Rayleigh number $Ra_{c}(Ta)$ is given as:
\begin{equation}
Ra_{c}(Ta) = \left[ (\pi^2 + k^2_{c})^3+\pi^2 Ta\right]/k^2_{c},
\label{Rc}
\end{equation}
where the critical wave number $k_{c}(Ta)$ satisfies the following equation:
\begin{equation}
2 \left( \frac{k_c^2}{\pi^2}\right)^3 + 3 \left( \frac{k_c^2}{\pi^2}\right)^2 = \left( 1 + \frac{Ta}{\pi^4} \right). \label{cubic_eq}  
\end{equation} 
Combining eqs.~\ref{cubic_eq} and ~\ref{Rc}, the critical Rayleigh number $Ra_c (Ta)$ may also be written as: 
\begin{equation}
Ra_c = 3 (\pi^2 + k_c^2)^2.\label{modified_Rc} 
\end{equation} 
The critical wavenumber, which is a real positive root of eq.~\ref{cubic_eq}, is: 
\begin{equation} 
k_{c}(Ta) = \pi \sqrt{a_{+} + a_{-} - 1/2},~\ ~\ ~
a_{\pm} = \left[ \frac{1}{4}\left\lbrace \frac{1}{2} + \frac{Ta}{\pi^4}\pm \sqrt{\left( \frac{1}{2} +\frac{Ta}{\pi^4}\right)^2 -\frac{1}{4}} \right\rbrace \right]^\frac{1}{3}.\label{kc}
\end{equation}

\subsubsection{Oscillatory convection}
The two growth rates $\sigma_{+}$ and $\sigma_{-}$ form a complex conjugate pair, if  $(k^2 Ra)^2 < 4 \pi^2 (\pi^2+k^2)^3 Ta$. Insertion of $\sigma_{\pm} = s \pm i \omega$ in equation~\ref{growth_rate} leads to the expressions for $s$ and $\omega$: 
\begin{equation}
s = \frac{-2({\pi}^2+k^2)^3 + k^2 Ra}{2({\pi}^2+k^2)^2}, ~~\ \omega (Ta) = \frac{\sqrt{4 {\pi}^2 (\pi^2+k^2)^3 Ta - {k^4}{Ra^2}}}{2({\pi}^2+k^2)^2}.\label{W0} 
\end{equation}
\noindent Setting $s=0$ leads to the condition for marginal stability  for oscillatory convection. The critical wavenumber $k_{\circ}$ and the critical Rayleigh number $Ra_{\circ}$ for the oscillatory instability in the $Pr \rightarrow 0$ limit are found to be $k_{\circ} = \pi/\sqrt{2} \approx 2.22$ and $Ra_{\circ} = 27 \pi^4/2 \approx 1315$, respectively. Notice that $k_{\circ}$ and $Ra_{\circ}$ are independent of $Ta$ in this limit. The angular frequency $\omega$ becomes real, if $Ta > 27 \pi^4/8 \approx 328$. The convection is oscillatory at the onset if $Ra_{\circ} < Ra_c (Ta)$. This is possible for $Ta \ge 548$, and the dimensionless angular frequency at the onset of oscillatory convection for $Ta = 548$ is 
$\omega_{\circ} = \sqrt{\frac{2}{3}\left( 548 - \frac{27 \pi^4}{8}\right)} \approx 12.1$. Simulations are done for parameters of the shaded region of the $Ra-Ta$ plane [see fig.~\ref{lin_stability} (a)]. The results of linear stability analysis are summarized in figure~\ref{lin_stability}. Blue triangles and circles in figure~\ref{lin_stability} (a) are values of $Ra_c (Ta)$ and $Ra_{\circ}$ respectively, computed from DNS. The fluid patterns near onset of oscillatory convection have been investigated in detail using DNS recently~\cite{pharasi_kumar_2013}. We have therefore carried out DNS in the vicinity of the bicritical point ($Ra_c = Ra_{\circ}$) and in the stationary convective regime. 

\section{Direct numerical simulations}
The direct numerical simulations (DNS) of the hydrodynamical system (equations~\ref{ns}-\ref{cont})  with the stress-free boundary conditions in the limit $Pr \rightarrow 0$ are carried out using the pseudo spectral method (see, Pharasi \& Kumar~\cite{pharasi_kumar_2013} for details) using an open source code Tarang~\cite{tarang}. 
All the components $v_{1}$, $v_{2}$ and $v_{3}$ of the velocity field are expanded as:
\begin{eqnarray}
v_1 (x,y,z,t) &=& \sum_{l,m,n} U_{lmn}(t) e^{i(lk_xx+mk_yy)} \cos{n\pi z},\label{eq.v1}\\
v_2 (x,y,z,t) &=& \sum_{l,m,n} V_{lmn}(t) e^{i(lk_xx+mk_yy)} \cos{n\pi z},\label{eq.v2}\\
v_3 (x,y,z,t) &=& \sum_{l,m,n} W_{lmn}(t) e^{i(lk_xx+mk_yy)} \sin{n\pi z}.\label{eq.v3}
\end{eqnarray}
\noindent  The perturbations with the critical wavenumber $k_c (Ta)$ are the most dangerous, as soon as the Rayleigh number $Ra$ is raised above the threshold $Ra_c (Ta)$ for convection. In addition, components with longer wavelengths (or smaller wave numbers) are always present in larger containers.  We have considered a rectangular simulation box of size $(2\pi/k_x)\times (2\pi/k_y) \times 1$. We can investigate various patterns arising from the nonlinear interaction of two wave vectors ${\bf k}_1$ and ${\bf k}_2$. The code allows various possibilities for the choice of fluid patterns. We may have $\mathbf{k}_1 = \alpha k_c \mathbf{e}_1 + \beta k_c \mathbf{e}_2$ and $\mathbf{k}_2 = \alpha k_c \mathbf{e}_1 - \beta k_c \mathbf{e}_2$ for a given choice of $\alpha$ and $\beta$. A nonlinear interaction of two wave vectors $\mathbf{k}_1$ and $\mathbf{k}_2$ generates a third wave vector $\mathbf{k}_3 =  -[\mathbf{k}_1 + \mathbf{k}_2]$. A pattern involving $\mathbf{k}_1$, $\mathbf{k}_2$ and $\mathbf{k}_3$ is excited, if they all fit in the chosen simulation box. The angle $\phi$ between two wave vectors $\mathbf{k}_1$ and $\mathbf{k}_2$ is then given by $\cos{\phi}$ $=$ ${\bf k}_1 \cdot {\bf k}_2/ |{\bf k}_1| |{\bf k}_2|$ $=$ $(\alpha^2 - \beta^2)/(\alpha^2 + \beta^2)$. The code can also investigate the interaction of two wave vectors ${\bf k}_1 = \alpha k_c \mathbf{e}_1$ and ${\bf k}_2 = \beta k_c \mathbf{e}_2$. The parameter $\alpha$ has been set equal to unity for most of the cases, while the parameter $\beta$ has been varied.  We allowed the ratio $\eta = k_x/k_y = \alpha/\beta$ to vary in the range $1/\sqrt{3} \le \eta \le 10$. The variation of the parameter $\eta$ has facilitated the investigation of a variety of fluid patterns. For $\eta \ge 4$, we observe the replacement of a set of wavy rolls by a new set of wavy rolls of different wave numbers oriented at an angle of $\phi = \arctan{(1/\eta)}$ at the primary instability for all possible values of $Ta$.  To investigate standard K\"{u}ppers-Lortz patterns, we set  $k_x = k_c/2$ and $k_y = \sqrt{3}k_c/2$.

The spatial grid resolutions of $64 \times 64 \times 64$ and $128 \times 128 \times 128$ have been used for DNS, which are good enough to resolve the flow structures at smaller rotation rates and smaller values of the reduced Rayleigh number $r = Ra/Ra_{c}(Ta)$. The time advancement is done using the standard fourth order Runge-Kutta (RK4) integration scheme. The time step $\delta t$ for the integration was $0.001$. We started a simulation for a given value of $Ta$ and $Ra$ with random initial conditions. The reduced Rayleigh number $r$ was increased in small steps ($0.001 \le \delta r \le  0.1$). The final values of all the fields after long time were used as initial conditions for the next run. We also did simulations at higher values of $r$ and decreased it in small steps to find out any hysteresis. We did not find any hysteresis in the entire range of parameters investigated in this paper. We performed several runs of the hydrodynamic system for a range of reduced Rayleigh numbers ($ 1 < r \le 1.25$) for different values of the ratio $\eta$. 

The horizontal aspect ratio $\eta$ of a simulation box does not necessarily mean the same of an experimental container in pseudo spectral method. The size of a rectangular simulation box fixes the lower cut-off values of the interacting wave vectors. A simulation box with square horizontal cross-section of side $L > 2\pi/k_c$ would allow interaction of two mutually perpendicular wave vectors of equal magnitude $k < k_c$.  A larger experimental container with a square horizontal cross-section allows several possibilities of an interaction between two wave vectors in mutually perpendicular directions. 

\begin{table*}[!]
\caption{Convective flows computed from the direct numerical simulation just above the primary instability ($r = 1.005$) for several values of Taylor number for boxes with different aspect ratio $\eta = k/q$. We have set $k_x = k_c (Ta)$ and $r = Ra/Ra_c$ for $0 \leq Ta \leq 550$ in DNS.  The flow patterns  show periodic bursting (PB) as well as chaotic bursting (CB) for $\eta = 1$, periodic wavy rolls (WR) at lower values of $Ta$ and K\"{u}ppers-Lortz (KL) patterns for $Ta \geq 40$ for $\eta = 1/\sqrt{3}$,  WR at smaller values of $Ta$ and KL and generalized  K\"{u}ppers-Lortz (GKL) patterns for higher values of $Ta$ for $\eta = 2$, and the KL patterns for $\eta \ge 4$, respectively.  We have set $k_x = k_o$ and $r = Ra/Ra_o$ for $Ta > 550$. We observe quasiperiodic cross-rolls in oscillatory regime ($Ta> 550$) for $\eta = 1$ and $2$, while KL or GKL patterns for $\eta = 1/\sqrt{3}$ and $\sqrt{3}$. The angles between the interacting rolls in the horizontal plane are also mentioned below the appropriate patterns. }
\begin{center}
\begin{tabular}{c|c|c|| ccccccccc}
\hline
$\alpha$ & $\beta$ & $\eta = \alpha/\beta $   & 	\multicolumn{6}{c}{Fluid Patterns at different values of Ta } \\
\cline{4-12}
 &   &   & Ta =$10$ & $25$ & $40$ & $50$ & $100$  & $500$ & $550$  & $575$  & $700$\\ 
\hline\hline
$1/2$ & $\sqrt{3}/2$ & 1/$\surd{3}$ & WR       & WR   & KL  & KL   & KL     & GKL    & GKL     & GKL     & KL\\

 & & & & & ($30.0^{\circ}$) & ($40.8^{\circ}$) & ($40.8^{\circ}$) & ($60.0^{\circ}$) & ($60.0^{\circ}$) & ($60.0^{\circ}$) & ($60.0^{\circ}$) \\
 & & & & & & & & ($30.0^{\circ}$) & ($30.0^{\circ}$) & ($30.0^{\circ}$) & \\
           
\hline
$1$ & $1$ & 1           & PB      & CB & PB & PB & CB    & CB  & CB   & QCR    & QCR\\
           
\hline
$1$ & $1/\sqrt{3}$ & $\surd{3}$  & WR        & WR   & WR   & WR   & KL & KL  & GKL & GKL & -\\
 & & & & & & & ($30.0^{\circ}$) & ($30.0^{\circ}$) & ($60.0^{\circ}$) & ($60.0^{\circ}$) & - \\
 & & & & & & & & & ($30.0^{\circ}$) & ($30.0^{\circ}$) & \\
            
\hline
$1$ & $1/2$ & 2           & WR        & KL  & KL  & KL  & KL     & GKL & GKL    & GKL & QCR\\

& & & & ($26.6^{\circ}$) & ($26.6^{\circ}$) & ($26.6^{\circ}$) & ($26.6^{\circ}$) & ($26.6^{\circ}$) & ($26.6^{\circ}$) & ($26.6^{\circ}$) & \\
& & & & & & & & ($63.4^{\circ}$) & ($63.4^{\circ}$) & ($63.4^{\circ}$) & \\
          
\hline
$1$ & $1/4$ & 4           & KL       & KL  & KL  & -    & -        & -  & - & - &   -\\ 
& & & ($14.0^{\circ}$) & ($14.0^{\circ}$) & ($14.0^{\circ}$) & & & & & & \\
           
\hline
$1$ & $1/10$ & 10	        & KL       & KL  & KL  & -    & -        &	& - & - &   - \\
& & & ($5.7^{\circ}$) & ($5.7^{\circ}$) & ($5.7^{\circ}$) & & & & & & \\
\hline           
 \end{tabular}
\label{table1}
 \end{center}
\end{table*}

\section{Results and discussions}
As the Rayleigh number $Ra$ is raised above a critical value $Ra_{c} (Ta)$ for a given value of $Ta$, the convection sets in. The fluid patterns at the instability onset depend, in general, on the Taylor number $Ta$ and the ratio $\eta$. Chaotic patterns were observed at the primary instability in certain windows of $Ta$ in all the simulation boxes. We observed periodic dynamics at the primary instability in smaller boxes ($\eta < 4$) for smaller values of $Ta$. The fluids patterns at the primary instability as computed from DNS are listed in Table~\ref{table1}.

\begin{figure}[h]
\centerline{\includegraphics[height=13 cm,width=14 cm]{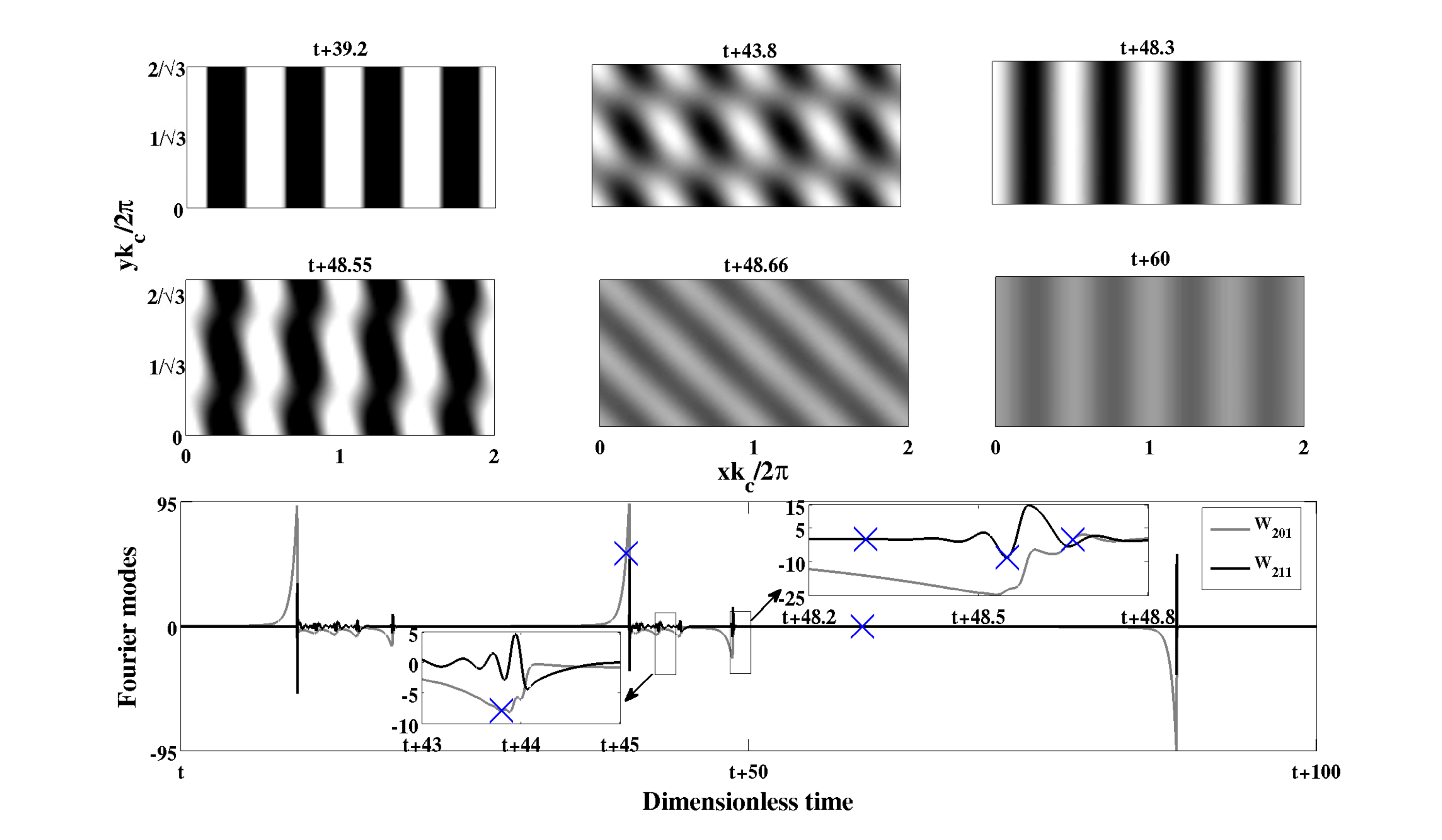}}
\caption[short]{Contour plots  of the temperature field at $z=0.5$ showing K\"{u}ppers-Lortz instability ($\eta = 1/\sqrt{3}$) at the onset ($r = 1.005$) for $Ta = 50$. The variations of the two largest Fourier modes $W_{201}$ and $W_{211}$ with time are shown in the lower row. The contour plots are for the time instants marked by cross ($ \times$) in the lower row. The intensity of the gray level in the contour plots is larger for larger value of the mode $|W_{201}|$.}\label{kl}
\end{figure}

\subsection{K\"{u}ppers-Lortz instability: $\eta = 1/\sqrt{3}$}
We now set $\alpha =1/2$ and $\beta = \sqrt{3}/2$, which fixes the parameter $\eta = 1/\sqrt{3}$. This allows the investigation of the interaction of two rolls in mutually perpendicular directions. In addition, it also allows us to investigate the interaction of two wave vectors $\mathbf{k}_1$ $=$ $\frac{k_c}{2}$ $(-\mathbf{e}_1 + \sqrt{3} \mathbf{e}_2)$ and $\mathbf{k}_2$ $=$ $-\frac{k_c}{2}$ $(\mathbf{e}_1 + \sqrt{3} \mathbf{e}_2)$.  The nonlinear interaction of two wave vectors in the latter case generates a third wave vector $\mathbf{k}_3$ $=$ $k_c \mathbf{e}_1$. The three wave vectors are of equal magnitude ($|\mathbf{k}_1|$ $=$  $|\mathbf{k}_2|$ $=$ $|\mathbf{k}_3|$ $=$ $k_c$), and they are oriented with each other at $120^{\circ}$. This is the well known case of K\"{u}ppers-Lortz instability~\cite{kl_1969, kuppers_1970}, who considered the limit $Pr \rightarrow \infty$ in their original work with stress-free boundaries. The corresponding patterns dynamics showed replacement of one set of straight rolls by another of set of rolls of the same wavelength but oriented with the older set at an angle of $60^{\circ}$. This situation is quite different than choosing ${\bf k}_1 = k_c \mathbf{e}_1$ and ${\bf k}_2 = \sqrt{3} k_c \mathbf{e}_2$, which gives $\eta = 1/\sqrt{3}$. This  describes an interaction of two wave vectors in mutually perpendicular directions: one with wave number equal to $k_c$ another with wavenumber equal to $\sqrt{3} k_c$. We have listed the values of $\alpha$ and $\beta$ in the first two columns of Table~\ref{table1}.

The replacement of a set of rolls by a new set of rolls of the same wavelength and inclined at an angle of $\phi$ with the old set of rolls is well known as K\"{u}ppers-Lortz (KL) patterns.  We observed in some cases that the wavelength of the new set of rolls was different from that of the old set. We have termed these patterns also as KL patterns here. The angle $\phi$ between two sets of rolls depends on the interacting wave vectors. We have also observed a new possibility. The first set of rolls is replaced by a second set of rolls oriented  at an angle $\phi$ with the first, and then the second set of rolls is replaced by a third set of rolls oriented at an angle $90^{\circ} - \phi$ with second set. This involves 
a competition of three sets of rolls of different wavelengths. These patterns are termed here as generalized K\"{u}ppers-Lortz (GKL) patterns. 

Clune and Knobloch~\cite{clune_knobloch_1993} investigated fluid patterns in rotating RBC using amplitude equations and found the possibility of standing waves at the instability onset at smaller rotation rates for $Pr < 0.442$ with no-slip boundary conditions. They showed that the critical value $Ta_{KL}$ of the Taylor number for KL instability  and the angle $\phi$ between two sets of oblique rolls of the same wavelengths at the onset of KL instability become smaller, as $Pr$ is decreased. They also predicted small angle instability for stress-free boundaries in case of a high Prandtl number fluid ($Pr = 100$). The angle between two wave vectors ${\bf k}_1 = \alpha k_c \mathbf{e}_1+ \beta k_c \mathbf{e}_2$ and ${\bf k}_2 = \alpha k_c \mathbf{e}_1- \beta k_c \mathbf{e}_2$ of the same magnitude $k_c$ would lead to small angle instability if $\alpha^2 + \beta^2$ and $\alpha^2 - \beta^2$ simultaneously tend to unity, which is not possible. The small angle instability involving two sets of rolls of the critical wavelength $\lambda_c$ is unlikely to occur unless $\beta \rightarrow 0$. The small angle instability, however, with any two wave vectors is possible in two cases: (i) $\alpha$ remains finite and $\beta \rightarrow 0$ or (ii) $\alpha \rightarrow 0$ and $\beta \rightarrow 0$. We do observe the angle $\phi$ decreasing as $\eta$ increases in the first case but there is no divergence observed in DNS for $\eta \le 10$. Clune and Knobloch also mentioned that some of their conclusions were valid for $Pr \sim O(1)$, and they would require modification in the limit of $Pr \rightarrow 0$.  

Cox and Mathews~\cite{cm_2000} investigated the instability of a set of straight rolls to a new set of similar rolls inclined at a small angle with the old set in case of stress-free boundary conditions using amplitude equations for $Pr \gg 1$. They predicted a system of rolls is always unstable to a new set of rolls at infinitesimal angle ($\phi \rightarrow 0$), if $Ta \ge Ta_c = 4 \pi^4$. However, Cox and Mathews considered only those modes for the vertical vorticity, which are independent of the vertical coordinates. This approach does not consider all possible nonlinear interactions. For example, the vorticity mode $Z_{211}$ ($Z_{121}$) interacts nonlinearly with the velocity mode $W_{101}$ ($W_{011}$) and generates other vorticity modes: $Z_{110}$, $Z_{112}$, $Z_{310}$ ($Z_{130}$) and $Z_{312}$ ($Z_{312}$). Similarly, the interaction of the vorticity mode $Z_{110}$ with the 2D roll mode $W_{101}$ ($W_{011}$) not only contributes to another 2D roll mode $W_{011}$ ($W_{101}$) but also generates higher velocity modes:  $W_{111}$,  and $W_{211}$ ($W_{121}$). We have observed that the  maxima of $|Z_{121}|$ and $|Z_{211}|$ are larger than the maxima of $|Z_{110}|$ in a finite size periodic simulation box, although the linear decay rate of the mode $Z_{110}$ is much smaller than the linear decay rates of the modes $Z_{211}$ and $Z_{121}$. The  infinitesimal angle instability in a small or narrow simulation box ($\eta \le 10$) is unlikely at least near onset. We have not found this instability for rectangular container with $\eta \le 2$ (please see patterns for $Ta \ge 500$ listed in Table~\ref{table1}). Any experimental container has a typical length scale in the horizontal plane, which restricts the maximum wavelength of the perturbations. Similarly, any periodic simulation box used in DNS using pseudo spectral method has a finite size. This imposes a maximum limit  on the possible wavelength of perturbations. This is independent of the fact whether one uses free-slip or no-slip boundaries on the top and bottom surfaces.  Although long wavelength perturbations are easily excited with free-slip boundaries, there is a maximum cut-off value in practice.  Square patterns were observed in  experiments with $CO_2$ ($Pr = 0.93$) and argon ($Pr = 0.69$) gases in a cylindrical container~\cite{bajaj_etal_1998}.  In addition, the bursting of patterns was not found in earlier studies~\cite{cm_2000, clune_knobloch_1993}  as observed in experiments by Bajaj et al~\cite{bajaj_etal_2002}. We have carried out DNS for the limiting case of $Pr \rightarrow 0$. The primary instability even without rotation is quite complex in very low-Prandtl-number fluids ($Pr << 1$) with stress-free boundaries~\cite{mishra_etal_2010, pal_etal_2013}. We have chosen  one side of the rectangular simulation box of a variable length and  the ratio $\eta = k_y/k_x$ is varied between $1/\sqrt{3}$ and $10$. Periodic simulation boxes with narrow rectangular cross-section or smaller square cross-section in the horizontal plane do not allow two sets of rolls at infinitesimal inclination.  

Various convective patterns computed from DNS for $\eta =1/\sqrt{3}$ at different values of $Ta$ are summarized in Table~\ref{table1}. The angle between the two sets of inclined rolls for KL or GKL patterns are mentioned in the brackets. A listing of two angles mean that old set of rolls may be replaced by a new sets inclined at any of the two values mentioned. We observed periodic wavy rolls at the primary instability for $Ta < 40$. KL instability is observed for $40 \le Ta \le 100$ and again for $Ta = 700$, while GKL patterns are observed for  $500 \le Ta \le 575$. GKL patterns are observed near the bicritical point. Figure~\ref{kl} shows contour plots of the temperature fields at $z=0.5$ just above the onset of convection ($r=1.005$) at six different instants for $Ta = 50$. The fluid patterns  consist of straight rolls, wavy rolls and oblique rolls, which appear irregularly in time. The temporal variations of the two largest Fourier modes $W_{201}$ (gray curve) and $W_{211}$ (black curve) corresponding to this sequence of patterns are displayed in the lower row of figure~\ref{kl}. The temporal variation of the leading modes shows unsteady convection at the primary instability. The contour plots are for the time instants marked by blue crosses in the third row of figure~\ref{kl}. Occasionally, one set of rolls are replaced by a new set of oblique rolls of the same wavelengths. However, there is a  difference from KL instability for $Pr \gg 1$. The leading Fourier mode $W_{201}$ describing a set of straight rolls keeps growing almost exponentially until it crosses a critical value. The wavy mode $W_{211}$ is then excited, and the straight rolls become wavy. Standing waves are generated along the roll axis. The generation of waves along the roll axis transfers the energy from the poloidal flow fields to the toroidal flow fields. The transferred energy dissipates quickly in viscous fluids. As a consequence, the magnitude of $W_{201}$ decreases sharply to a very small value. The waves along the rolls disappear a little later. The mode $W_{201}$ again starts growing and the waves are generated only when roll intensity crosses a critical value. When $W_{211}$ is comparable to $W_{201}$, the straight or wavy rolls are replaced by a new set of rolls oriented at an angle  $\phi \approx 40^{\circ}$. The bursting of patterns has been observed in experiments by Bajaj et al~\cite{bajaj_etal_2002}. The variation of the amplitude of the roll modes also showed critical bursting.  This kind of behaviour is also observed in a low-dimensional model of convection in very low-Prandtl-number fluids~\cite{kumar_etal_2006}. The phenomenon of critical bursting is known to occur in the spring-block model~\cite{carlson_langer_1989}. KL instability shows the features of the critical bursting in the limit $Pr \rightarrow 0$. 

\subsection{Patterns in a small square box: $\eta = 1$}

\begin{figure}[h]
\centerline{\includegraphics[height=12 cm,width=14 cm]{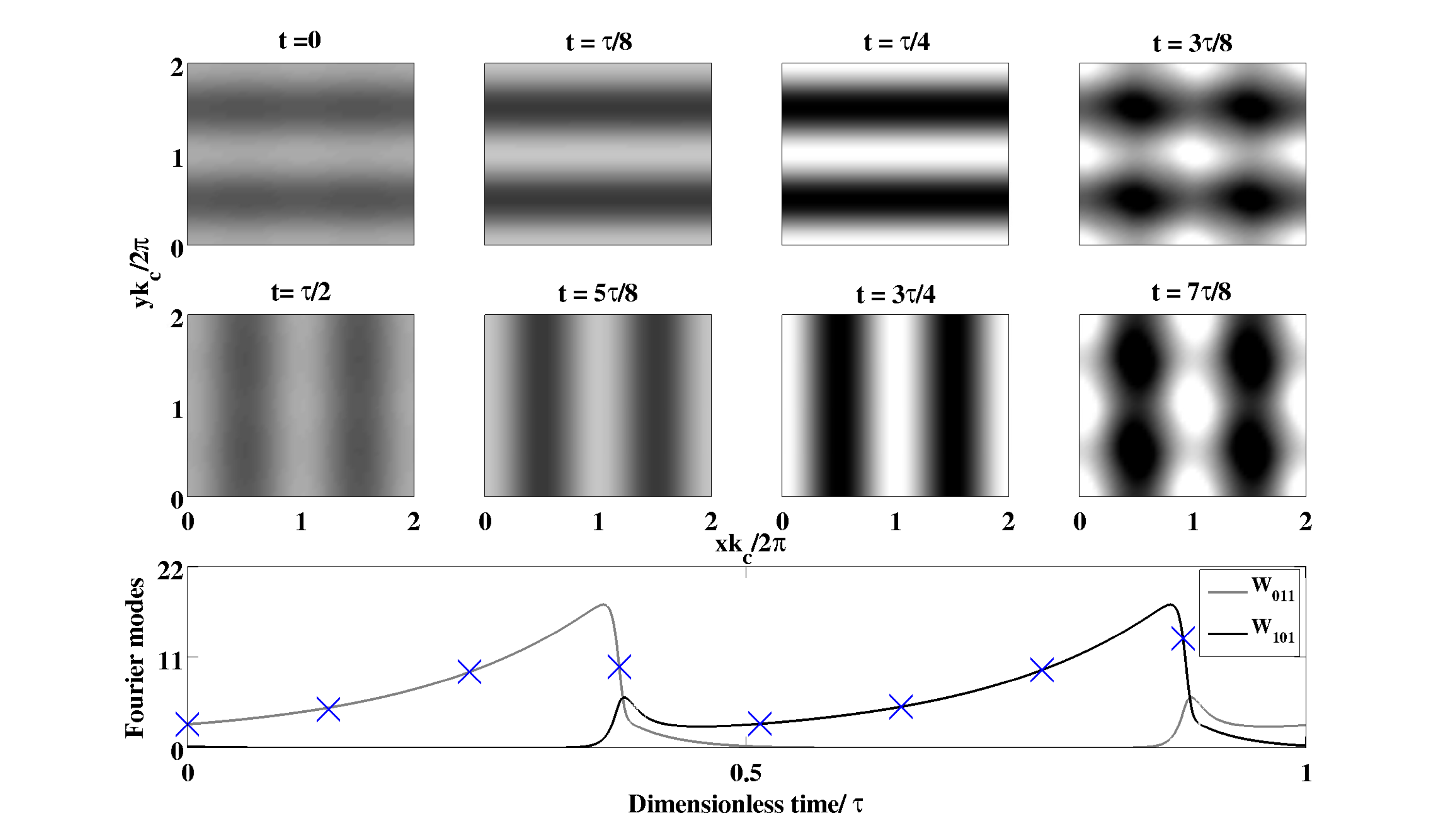}}
\caption[short]{Periodic competition of two sets of mutually perpendicular  straight rolls for $r = 1.001$ and $Ta=10$, as obtained from DNS in a small square box [$k_x = k_y = k_c (Ta)$, $\eta =1$]. The mid-plane ($z = 0.5$) contour plots of the temperature field are shown at eight equally spaces instants during the oscillation period ($\tau = 12.74$) in the first two rows. The black (white) regions in the contour plot correspond to relatively cooler (warmer) fluid. The third row displays the temporal variation of the two largest modes $W_{101}$ (black curve) and $W_{011}$ (gray curve).}
\label{sq_box}
\end{figure}

The fluid patterns are quite different in a simulation box with a square horizontal cross-section ($\eta = 1$). The simulations show a competition between two sets of mutually perpendicular straight rolls at the primary instability. The dynamics may be periodic or chaotic depending upon the value of $Ta$. The evolution of convective patterns with time is displayed in the first and second rows of figure~\ref{sq_box} for $r = 1.001$ and $Ta = 10$. Patterns are displayed at eight equally spaced instants during one period ($\tau$) of oscillation. The patterns consist of a set of straight rolls either parallel to the $x$ axis or parallel to the $y$ axis for more than  two-thirds of the  oscillation period. The change in orientation of rolls occurs within a very short period. The temporal variation of the leading Fourier modes $W_{101}$ (black curve) and $W_{011}$ (gray curve) for these patterns is shown in the third row of figure~\ref{sq_box} for one oscillation period. The Fourier mode $W_{011}$ grows exponentially with time, while the other mode $W_{101}$ decays to zero.  The mode $W_{101}$ is excited from its zero value when the growing $W_{011}$ exceeds a critical value, which is slightly less than its maximum value. This stops further growth of the rolls parallel to the $x$ axis. The mode $W_{011}$ starts decaying sharply, while the mode  $W_{101}$ keeps growing. We observe square patterns when $|W_{011}| = |W_{101}|$ and cross-rolls for $|W_{011}| \neq |W_{101}|$. These cross-rolls are also known as asymmetric squares, as $k_x = k_y = k_c$. In a very short period, the rolls parallel to the $x$ axis disappear. We then observe rolls growing parallel to the $y$ axis. A set of rolls parallel to the $x$ axis is once again excited, as soon as the fluid velocity in the set of rolls parallel to the $y-$ axis exceeds a large ($\gg 1$) critical value. This is also an example of critical bursting of rolls. Patterns show periodic bursts (PB) as well as chaotic bursts (CB) as $Ta$ is varied. The period of exponential growth of any set of rolls to a critical value is much larger than the  period of bursting. The phenomenon of bursting reported here is an example of non-local bifurcation at the instability onset. This behaviour cannot be captured in an analysis based on local bifurcation theory around the conduction state or a stationary convective state. These patterns were not reported before at the instability onset in a rotating RBC. They are qualitatively new and more likely to occur in a square container. The Fourier modes $W_{101}$ and $W_{011}$ show relaxation oscillation due to the presence of two time scales. Their temporal  evolution of these modes have strong resemblance with the time series recorded by Bajaj et al.~\cite{bajaj_etal_2002} leading to bursting of patterns. 

\begin{figure}[h]
 \begin{center}
\includegraphics[height=10 cm, width=14 cm]{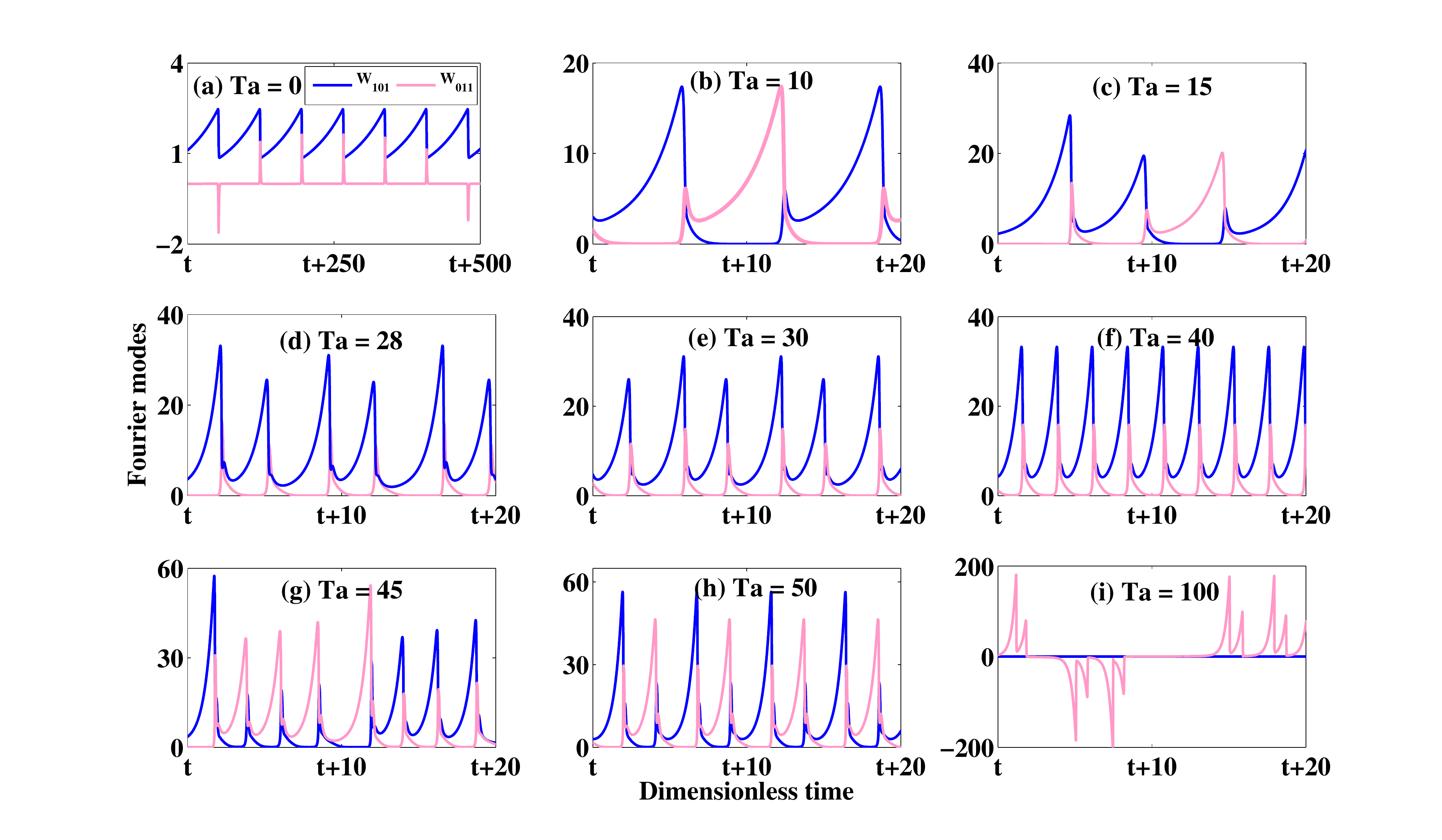}
\caption[short]{Temporal variation of the two largest Fourier modes $W_{101}$ (blue curves) and $W_{011}$ (pink curves) near primary instability ($r= 1.001$) for different values of $Ta$ for $\eta = 1$. As soon as the mode $W_{101}$ ($W_{011}$) reaches a critical value, the mode $W_{011}$ ($W_{101}$) gets excited.  The mode $W_{101}$ ($W_{011}$) decays very sharply to zero, while the other mode keeps growing.  For $Ta < 20$, both the modes alternately decay to zero. For $Ta > 20$, one of the modes decays to zero while the other mode falls to a finite small value. Both sets of rolls show the phenomenon of critical bursting.}\label{sq_modes}
\end{center}
\end{figure}
 
\begin{figure}[h]
 \begin{center}
\includegraphics[height=10 cm, width=14 cm]{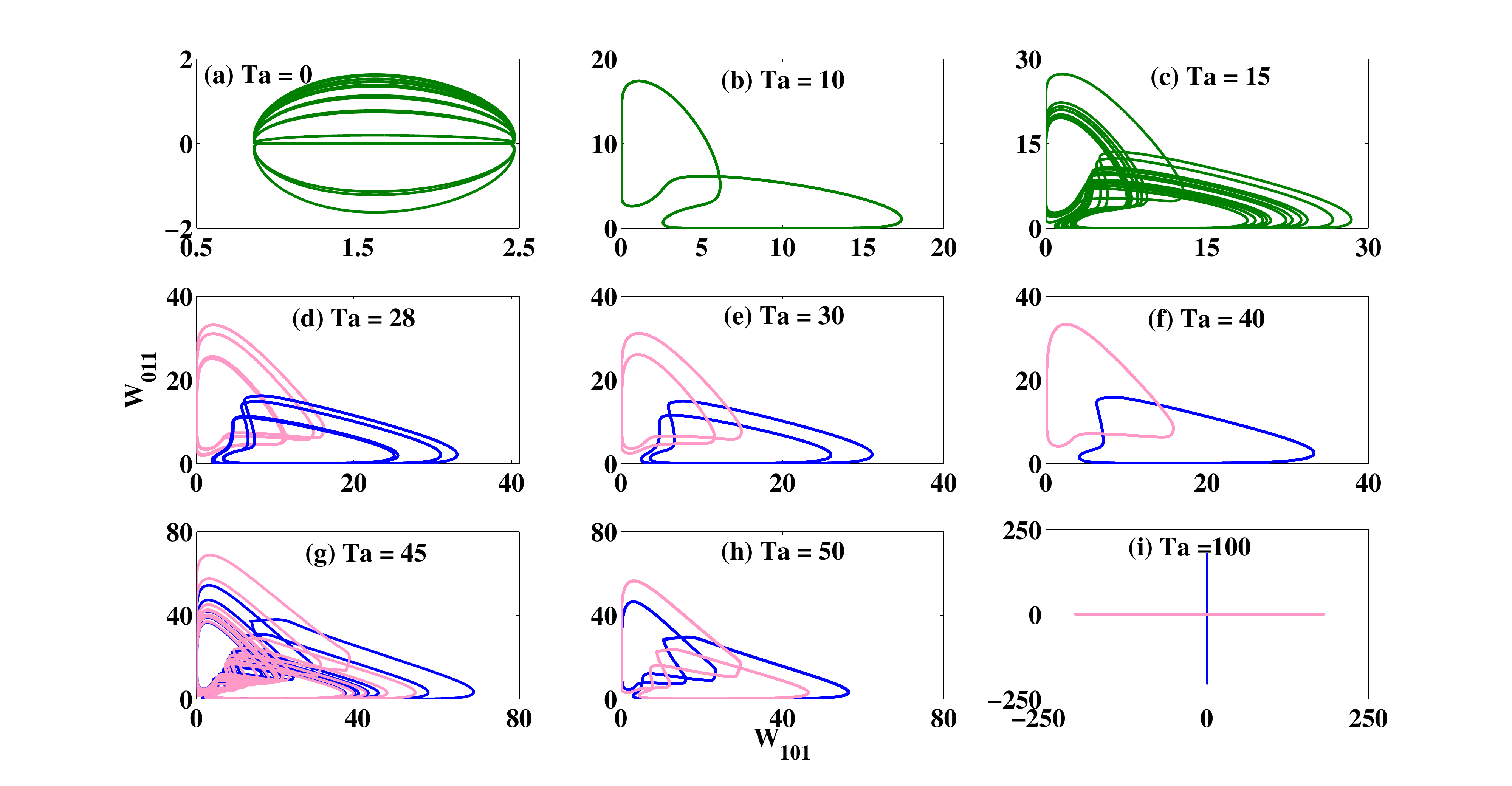}
\caption[short]{The projection of the phase portraits on the  $W_{101}-W_{011}$ plane near the primary instability ($r= 1.001$) for different values of $Ta$ ($\eta = 1$). Two sets of possible phase plots (blue and pink curves) are possible in each quadrant of the $W_{101}-W_{011}$ plane. The green curves show the glued solutions. The temporal variation of these modes are given in figure~\ref{sq_modes}.}
\label{sq_phase_plots}
 \end{center}
\end{figure} 

The temporal evolutions of the two largest Fourier modes $W_{101}$ (blue curves) and $W_{011}$ (pink curves) slightly above the onset of convection ($r=1.001$) are shown in figure~\ref{sq_modes} for $\eta = 1$ at different values of $Ta$. In the absence of rotation ($Ta = 0$), the solution is always chaotic at the onset~\cite{pal_etal_2009, mishra_etal_2010} in the limit $Pr \rightarrow 0$~[see figure~\ref{sq_modes} (a)]. The two largest Fourier modes $W_{101}$ and $W_{011}$ vary chaotically in time except in narrow windows of $Ta$ near $Ta = 10$ and $Ta = 40$.  Figure~\ref{sq_phase_plots} shows the phase portraits of the flow dynamics in the  $W_{101}-W_{011}$ plane corresponding to the temporal evolution of the Fourier modes given in Figure~\ref{sq_modes}. Unsteady solutions in the different quadrants of the $W_{101}-W_{011}$ plane get entangled at the onset for $Ta = 0$ [see figure~\ref{sq_phase_plots} (a)]. The glued periodic and chaotic solutions are observed only in one of the four quadrants of the $W_{101}-W_{011}$ plane for $Ta \neq 0$ [see figure~\ref{sq_phase_plots} (b)-(i)]. A glued limit cycle at smaller values of $Ta$ become chaotic as $Ta$ is raised. The chaotic glued oscillation breaks into two sets of distinct solutions with further increase in $Ta$. Figure~\ref{sq_phase_plots} (d), (e) and (f) show  two distinct sets of period-four, period-two and period-one limit cycles at $Ta = 28$, $30$ and $40$ respectively. The fluid patterns appeared as cross-rolls due to the competition between two sets of parallel rolls in mutually perpendicular directions. The competition was found to be periodic as well as chaotic with the variation of $Ta$.  A transition from chaotic cross-rolls to chaotic rolls was observed for larger values of $Ta$ [see figure~\ref{sq_modes} (i)]. The temporal variations of both the Fourier modes show critical bursting. The bifurcation at the primary instability is nonlocal for  $Ta \le 100$. The second row of Table~\ref{table1} lists various possible fluid patterns computed from DNS for $\eta =1$. 

\begin{figure}[h]
\begin{center}
\includegraphics[height=9.5 cm, width=13.0 cm]{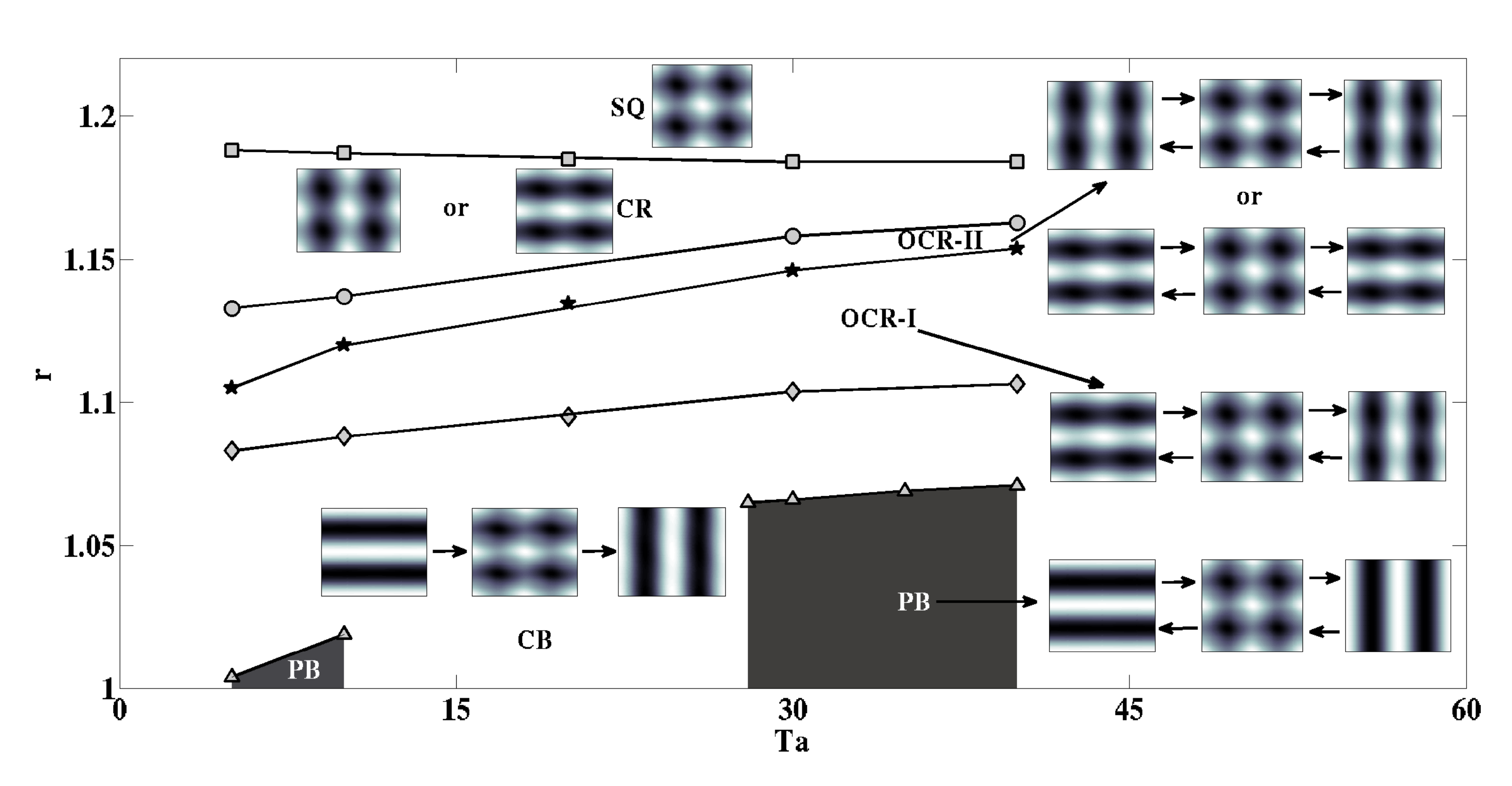}
\caption[short]{Regions of various flow patterns in the $r-Ta$ plane as obtained from DNS: The boundary of periodic bursts (PB) and chaotic bursts (CB) of two sets of rolls is marked by `$\triangle$', while that of CB solutions and a glued oscillation of two sets of cross-rolls (OCR-I) is marked by `$\lozenge$'. The boundary of two types of  oscillatory cross-rolls (OCR-I \& OCR-II) is marked by `$\bigstar$', while that of the OCR-II and the stationary cross rolls (CR) is marked by `$\circ$'. The square (SQ) patterns are observed above the boundary marked by `$\Box$'. The primary instability is periodic in windows of Taylor number $2 \le Ta \le 10$ and $28 \le Ta \le 40$, but chaotic for $10 < Ta < 28$. The corresponding patterns are shown in the inset of the diagram.}
\label{r_Ta_space}
\end{center}
\end{figure}

\begin{figure}[h]
\centerline{\includegraphics[height=6 cm,width=12 cm]{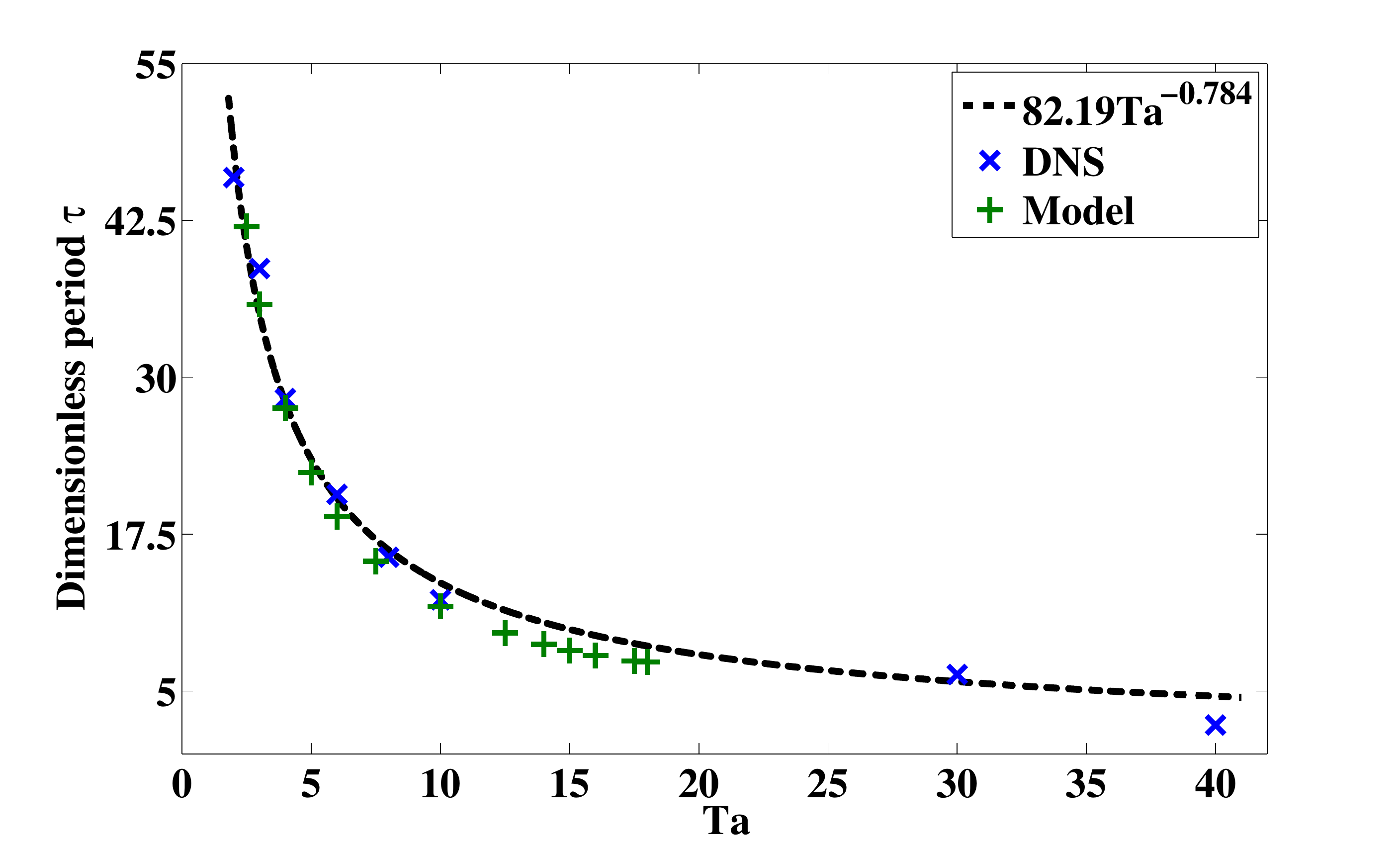}}
\caption[short]{Variation of the dimensionless time period $\tau$ of self-tuned oscillation (PB) at primary instability ($r=1.001$) as a function of Taylor number $Ta$. The time period $\tau$ increases with decreasing Taylor number $Ta$. The best fit (the dashed curve) to the data obtained by DNS (marked as `$\times$') for $2 < Ta \le 40$ shows  $\tau \sim Ta^{-0.784}$. The values of $\tau$ computed from the model (shown by the symbol `+') agree well with those computed from DNS.}
\label{period_limit_cycles}
\end{figure}

\begin{table*}[h]
\caption{Flow patterns for different values of the reduced Rayleigh number $r$ and the Taylor number $Ta$ observed in a simulation box with $\eta = 1$. The patterns are: periodic bursting (PB), chaotic bursting (CB), oscillating cross-rolls (OCR-I, $|W_{101}|_{max} = |W_{011}|_{max}$) but with a finite phase difference between them, oscillating cross-rolls (OCR-II, $|W_{101}|_{max} \neq |W_{011}|_{max}$), stationary cross-rolls (CR) and stationary squares (SQ).}
\begin{center}
  \begin{tabular}{ccccc}
\hline
 Fluid patterns  &r($Ta = 5$) &\multicolumn{2}{c}{r($Ta=10$)}& r($Ta=30$) \\
\cline {3-4}
  &DNS&  DNS & Model & DNS\\ 
\hline\hline
PB	&1.001 - 1.003~& 1.001  -1.018~& 1.001 -  1.05~& 1.001 - 1.065 \\
\hline
CB	&1.004 - 1.082~& 1.019 - 1.088~&       -      ~& 1.066 - 1.103\\
\hline
OCR-I	&1.083 - 1.105~& 1.089 - 1.120 ~& 1.051 - 1.137~& 1.104 - 1.146\\
\hline
OCR-II	&1.106 - 1.132~& 1.121 - 1.137~& 1.137 -  1.157~& 1.147 - 1.158\\
\hline
CR	&1.133 - 1.188~& 1.138 - 1.187~& 1.158 -  1.194~& 1.159 - 1.184\\
\hline
SQ	& $\leq 1.189$~& 1.188  - 1.250~& 1.195 - 1.250~& 1.185 - 1.250 \\
\hline
 \end{tabular}
\label{table2}
 \end{center}
\end{table*}

Various Fluid patterns, computed from DNS for different regions of the parameter space $r - Ta$, are plotted in figure~\ref{r_Ta_space} for $\eta = 1$. Two sets of straight rolls are found to alternate at the primary instability in a square box. The dynamics is observed to be periodic (PB) for $2 \le Ta \le 10$ (shaded regions in figure~\ref{r_Ta_space}) and $28 \le Ta \le 40$, while it is found to be chaotic (CB) for $10 < Ta < 28$. The two regimes are found to alternate  as $Ta$ is increased further. The primary instability becomes chaotic for $Ta > 55$.  Boundaries of regions in the $r-Ta$ plane separating different convective patterns are marked by different symbols. The boundary of the PB and the CB patterns is marked by `$\triangle$' in figure~\ref{r_Ta_space}. As the reduced Rayleigh number $r$ is raised for a fixed value of $Ta$ ($< 40$), the minimum of the mode $W_{101}$ or $W_{011}$ becomes finite. We then observe a competition between straight rolls and cross-rolls. The minima of both the rolls become finite with further increase in $r$, and the corresponding patterns are oscillatory cross-rolls.  We observe two possibilities: (i) two sets of rolls have equal amplitude but a phase difference between them. They appear as square patterns only for certain instants when $W_{011} (t) = \pm W_{101} (t)$, and as cross-rolls otherwise. They represent a periodic competition between two sets of oscillating cross-rolls. We have labeled them as OCR-I patterns.  (ii) Two sets of rolls have unequal amplitudes. As two different solutions are possible ($|W_{101}| > |W_{011}|$ or $|W_{101}| < |W_{011}|$), we have labeled them as OCR-II. The boundary separating CB patterns from  OCR-I patterns is marked by `$\lozenge$', and that separating OCR-I and OCR-II is marked by `$\bigstar$'. If $r$ is raised above the boundary marked by `$\circ$', we observe a transition from oscillating cross-rolls (OCR-II) to stationary cross rolls (CR) via an inverse Hopf bifurcation. The range of $r$ for which we observe stationary cross-rolls decreases with increasing $Ta$. We also observe two sets of stationary cross- rolls corresponding to two sets of oscillating cross-rolls. Raising $r$ above the boundary marked by `$\Box$' leads to a regime where the stationary square patterns are observed. The transition from two sets of stationary cross-rolls to stationary square patterns occurs via a backward pitchfork bifurcation. We observed stationary square patterns at moderate rotation rates for $r \ge 1.19$. Bajaj et al~\cite{bajaj_etal_1998} observed square patterns in a cylindrical container ($\eta = 1$) at higher rotation rates in water ($Pr = 5.4$) at $r = 1.09$  and in argon ($Pr = 0.69$) at $r = 1.04$. 

Table~\ref{table2} lists the various flow patterns for different values $r$ at three different rotation rates ($Ta$ $=$ $5$, $10$ and $30$). The patterns labeled as PB and CB cannot be captured in a local analysis around the conduction state. The instabilities leading to PB and CB states are therefore non-local. The variation of dimensionless time period $\tau$ of the periodic bursting (PB) as a function of Taylor number $Ta$ is shown in figure~\ref{period_limit_cycles}. The points marked by `$\times$' are computed from the DNS. The points marked by `$+$' are obtained from a model to be introduced in the next section. The time period $\tau$ of PB increases monotonically with decreasing Taylor number $Ta$ for $2 < Ta \le 40$. The convective dynamics is chaotic for $Ta \le 2$, and is discussed in detail in Sec.~\ref{model2}.    

\begin{figure}[h]
 \begin{center}
\includegraphics[height= 10 cm, width=14 cm]{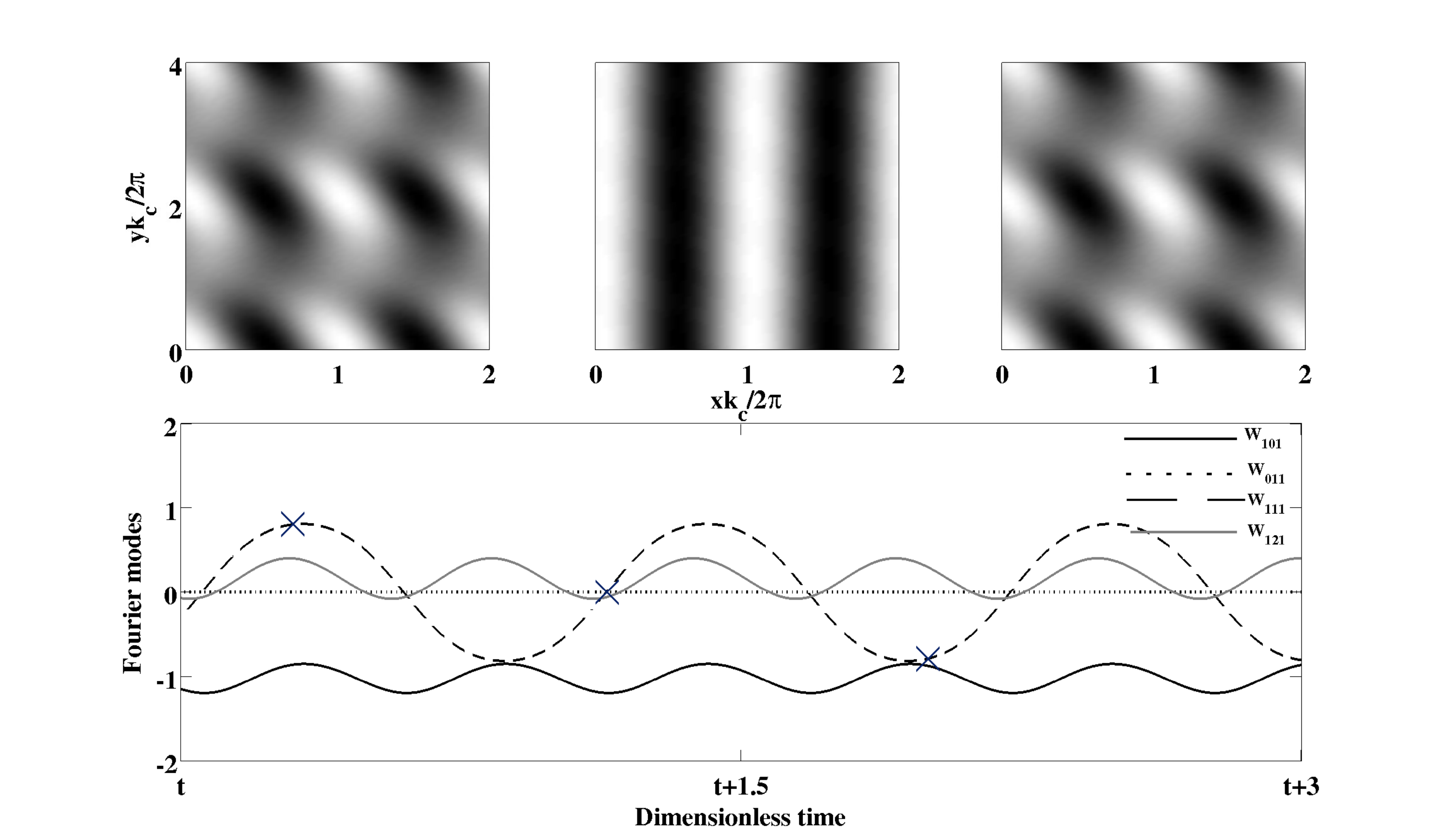}
\caption[short]{Contour plots of the temperature field, computed from DNS for $z =0.5$, $Ta = 10$ and $r = 1.005$ in a rectangular periodic box ($\eta = 2$) for three time instants are shown in the upper row. The temporal variation for Fourier modes $W_{101}$ (solid black curve), $W_{011}$ (dotted black curve), $W_{111}$ (dashed black curve), and $W_{121}$ (solid gray curve) are displayed in the lower row. The convective patterns (left to right) are computed at time instants marked by `$\times$' (left to right) in the lower row. The plots show a periodic competition between  wavy and straight rolls.}
\label{rect_box_small}
 \end{center}
\end{figure} 

\begin{figure}[h]
 \begin{center}
\includegraphics[height=14 cm, width=14 cm]{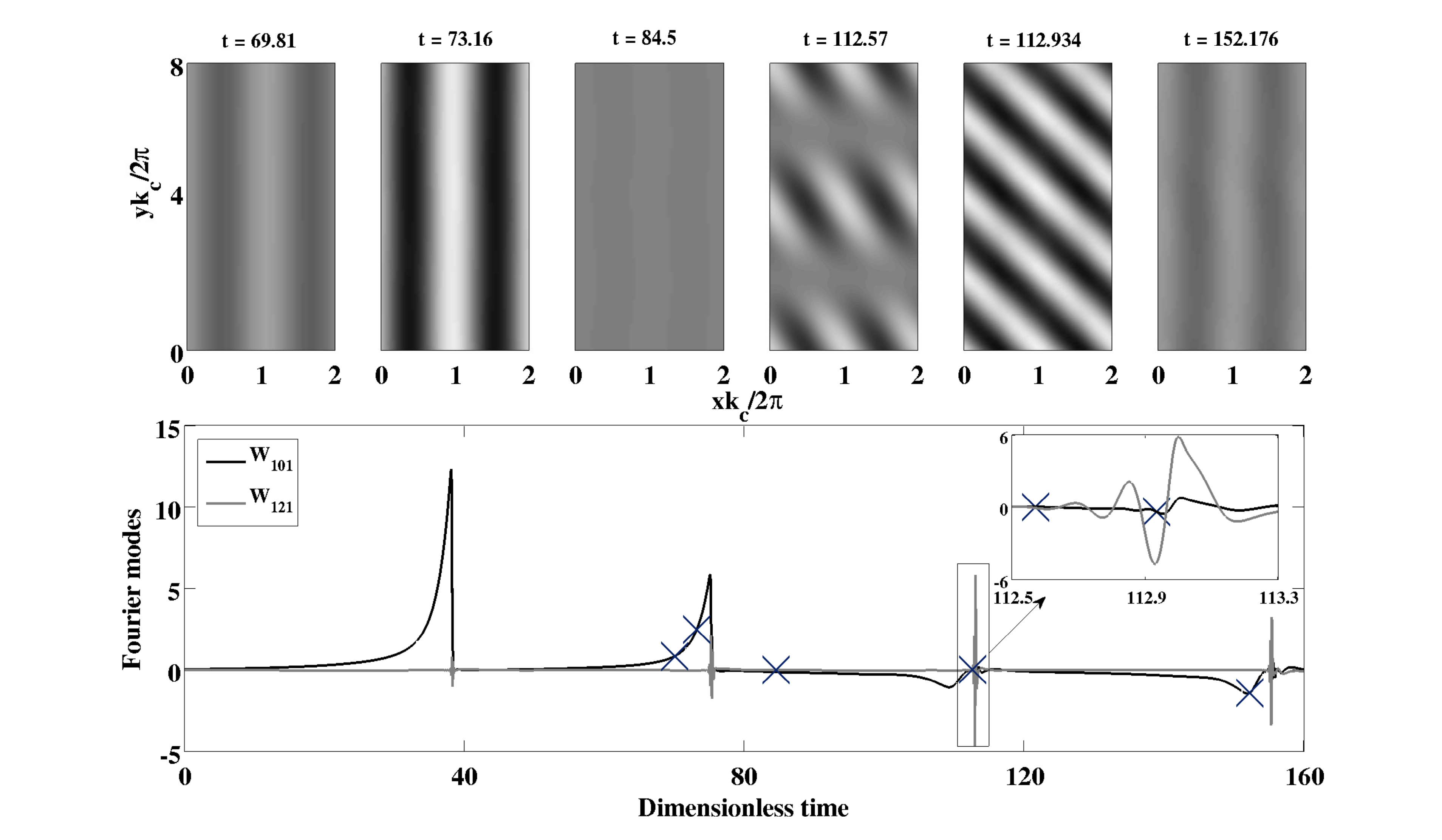}
\caption[short]{Chaotic competition between two sets of rolls inclined at an angle of $\phi = 14^{\circ}$ for $\eta = 4$ and $Ta = 10$ just above the onset ($r = 1.005$) of convection. The upper row shows the contour plots of temperature field at $z = 0.5$. The temporal variation of two large-scale Fourier modes $W_{101}$ (black curve) and $W_{121}$ (gray curve) is plotted in the lower row. Six contour plots are computed at time instants marked by `$\times$' on the curve shown in the lower row.}
\label{rect_box_large}
 \end{center}
\end{figure}

\subsection{Patterns in larger boxes: $\eta \ge 2$}

The convective patterns in a rectangular  box ($\eta = 2$) are shown in figure~\ref{rect_box_small}. The contour plots of the temperature field at $z=0.5$ for three time instants are displayed in the upper row. The fluid patterns are time-periodic wavy rolls. The lower row of figure~\ref{rect_box_small} shows the temporal variations of four velocity modes: $W_{101}$ (solid black curve), $W_{011}$ (dotted black curve), $W_{111}$ (dashed black curve), and $W_{121}$ (solid gray curve). The mode $W_{011}$  always remains  zero. The other three modes vary periodically in time. The  modes $W_{101}$ and $W_{121}$ have the same temporal period, but a small phase difference.  The time period of the mode $W_{111}$ is twice that of $W_{101}$. The 2D roll mode $W_{101}$ and the nonlinear mode $W_{121}$ have non-zero mean, but $|W_{101}|$ $>$ $|W_{121}|$. The nonlinear mode $W_{111}$ has zero mean. The convective patterns are periodic standing waves along the roll-axis. We have labeled these patterns as wavy rolls (WR). Whenever $W_{111} = 0$ and $|W_{101}|$ $\gg$ $|W_{121}|$, the patterns appear as almost 2D rolls. The sequence of patterns is quite different from those observed for $\eta = 1$ and does not involve any critical amplitude of a set of growing rolls to trigger waves along the rolls.  

The temporal evolution of convective patterns near the primary instability ($r = 1.005$) in a larger simulation box ($\eta=4$) for $Ta =10$, as obtained from DNS, is displayed in figure~\ref{rect_box_large}. The contour plots of the temperature field $\theta$ at $z = 0.5$ are shown in the upper row. The convective patterns consist of straight (2D) rolls, wavy rolls and oblique wavy rolls. Straight rolls with different intensities are observed for most of the time. The flat surface at $t = 84.5$ represents the state after bursting of rolls. Three-dimensional (3D) patterns appear only for a very short period. The temporal variations of the two largest velocity modes $W_{101}$ (black curve) and $W_{121}$ (gray curve) for this sequence of patterns are shown in the lower row of figure~\ref{rect_box_large}. Both modes vary chaotically with time. The least absolute value of the  mode $W_{101}$ is very small ($\sim 10^{-2}$), while that for the mode $W_{121}$ is almost zero. The 2D roll mode $W_{101}$ initially grows exponentially as predicted by the linear theory, while the nonlinear mode $W_{121}$ remains practically zero. As soon as  $W_{101}$ exceeds a large critical value, the nonlinear mode $W_{121}$ gets excited, as in the case of $\eta = 1$ and $1/\sqrt{3}$. We observed GKL patterns for $\eta \ge 4$. The oblique wavy rolls appear due to the generation of the wavy modes (e.g. $W_{121}$), which are expected in low-Prandtl-number fluids. The angle $\phi$ between the axis of straight rolls and that of the oblique rolls is equal to $14^{\circ}$ in this case.  For larger simulation boxes ($2 < \eta \le 10$), the qualitative behaviour is similar. The angle $\phi$, given by the relation $\phi = \arctan{(k_x/k_y)} = \arctan{(1/\eta)}$, decreases as $\eta$ increases. As $\eta \rightarrow \infty$ (i.e., $k_y \rightarrow 0$), the angle $\phi \rightarrow 0$. DNS with stress-free boundaries on a periodic rectangular lattice with $\eta \le 10$ rules out the divergence due to small angle instability.  Oblique rolls and a chaotic competition between two sets of mutually perpendicular rolls were observed in DNS~\cite{cb_2000} for rotating convection with no-slip boundary conditions. 

\begin{figure}[h]
 \begin{center}
\includegraphics[height=10 cm, width=14 cm]{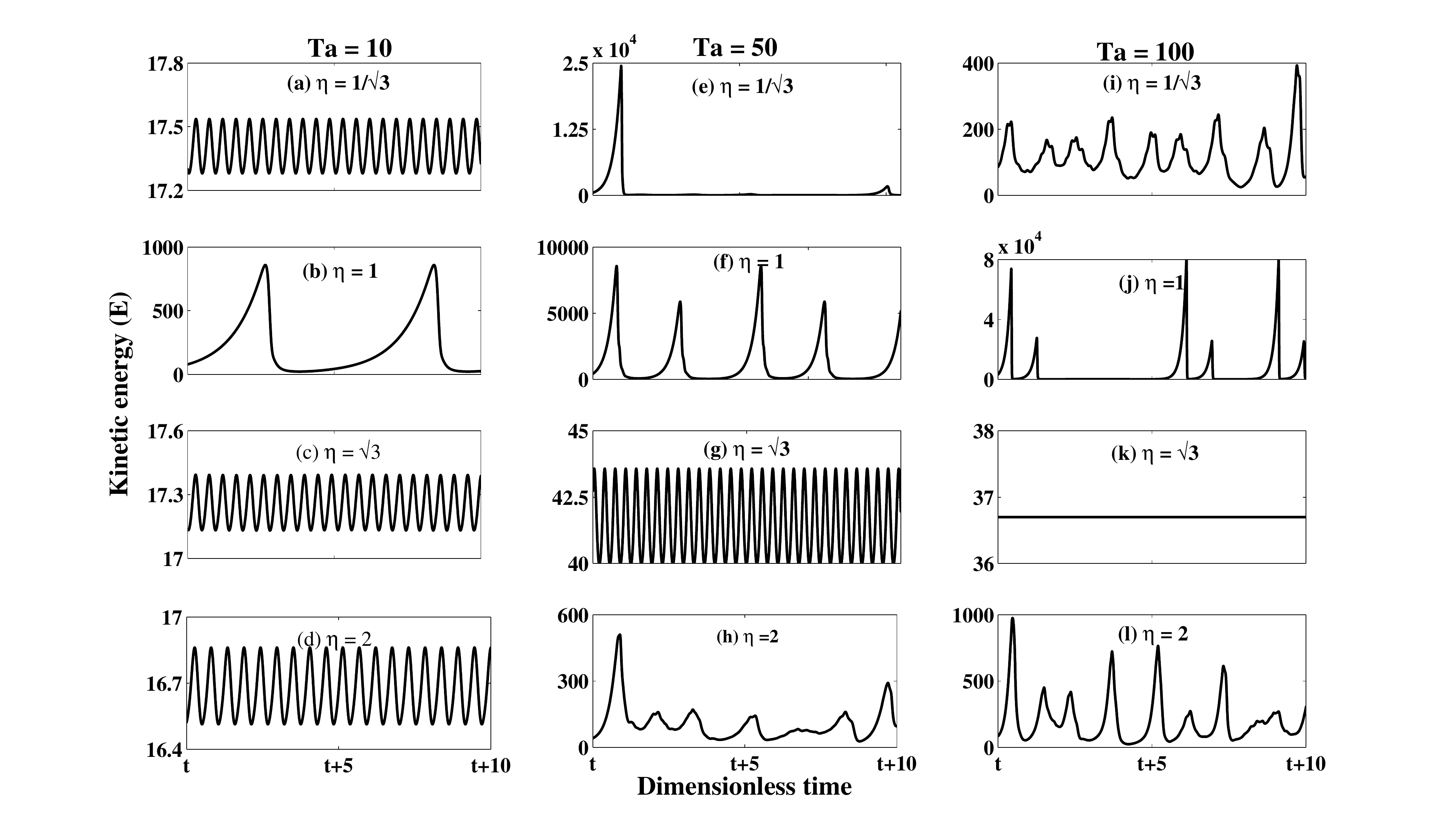}
\caption[short]{Temporal variation of the boxed averaged kinetic energy (E) for different values of $\eta$ and $Ta$ near the onset of convection ($r = 1.005$).}
\label{energy_time}
 \end{center}
\end{figure}

\subsection{Kinetic energy of fluid patterns}
Figure~\ref{energy_time} displays a comparison of the spatially averaged kinetic energy $E$ of different patterns near onset ($r = 1.005$) for four different values of the horizontal aspect ratio and three different values of the rotation rate. The comparison of $E$ for different values of $\eta$ for $Ta=10$ is given in the first column. Although all the patterns are time periodic, the kinetic energy for $\eta =1$ is $50$ times more than for other values of $\eta$. For 
$Ta = 50$ (the second column), the convection is always chaotic except $\eta = \sqrt{3}$. The maximum kinetic energy is observed for KL patterns, but PB yields maximum value of the time and space averaged kinetic energy $\bar{E}$.  For $Ta = 100$ (the third column), the energy is higher for CB patterns than GKL and KL patterns. For a given value of $\eta$, $E$ at the primary instability increases with the rotation rate. In addition, the convection is likely to be chaotic at the onset in very small-Prandtl-number fluids at 
moderate and high values of $Ta$.

\section{Model-I: A low dimensional model for $\eta = 1$}\label{model1}
Direct numerical simulations are hugely time consuming for computing the bifurcation diagram. Amplitude equations~\cite{cm_2000, dawes_2001a, dawes_2000} are usually constructed in the case of a local bifurcation to save time and understand unfolding of bifurcations better. Amplitude equations assuming local bifurcations at the primary instability may not be suitable to capture the fluid patterns in zero-Prandtl-number convection with rotation. We therefore construct a low-dimensional dynamical system to capture the sequence of bifurcations in zero-Prandtl-number convection with rotation. We first eliminate the pressure term from hydrodynamic equations by applying curl ($\boldsymbol{\nabla \times}$) once to Eq~\ref{ns}. This leads to an equation for the vorticity $\boldsymbol{\omega} = {\boldsymbol{\nabla \times v}}$, whose vertical component $\omega_{3}$ is given by, 
\begin{equation}             
\partial_{t}\omega_{3} = \nabla^{2}\omega_{3} + \sqrt{Ta}\partial_{z}v_{3} + 
[(\boldsymbol{\omega\cdot\nabla})v_{3} - (\boldsymbol{v\cdot\nabla})\omega_{3}]. \label{vorticity_model}   
\end{equation}
Operating by curl twice ($\boldsymbol{\nabla \times \nabla \times}$) on Eq~\ref{ns} and using Eq~\ref{cont} leads to an equation for the vertical velocity  $v_{3}$ given as:
\begin{equation}
\partial_{t}(\nabla^{2} v_{3}) = \nabla^{4} v_{3} + Ra{\nabla^{2}_{H}}\theta - \sqrt{Ta}\partial_{z}\omega_{3}
-\boldsymbol{\lambda}\boldsymbol{\cdot\nabla\times} [(\boldsymbol{\omega\cdot\nabla})\boldsymbol{v} - (\boldsymbol{v\cdot\nabla})\boldsymbol{\omega})], \label{velocity_model}
\end{equation}
We expand $v_3 (x, y, z, t)$ and $\omega_3 (x, y, z, t)$ in the Fourier modes as:
\begin{center}
\begin{eqnarray}
{v_{3}} &=& \sum_{l,m,n} [\tilde{W}_{lmn}(t)\cos{lkx}\cos{mky} 
+ \tilde{W}_{\bar{l}\bar{m}n} (t)\sin{lkx} \sin{mky}]\sin{n{\pi}z}, \label{eq.vel_model}\\
{\omega_{3}} &=& \sum_{l,m,n}[\tilde{Z}_{lmn}(t)\cos{lkx}\cos{mky} 
+ \tilde{Z}_{\bar{l}\bar{m}n}(t)\sin{lkx}\sin{mky}]\cos{n{\pi}z}. \label{eq.vort_model}
\end{eqnarray}
\end{center}
We set $k = k_c (Ta)$ in the model, as is done in DNS. The horizontal velocities $v_{1}$, $v_{2}$ and the horizontal vorticities $\omega_{1}$, $\omega_{2}$ may be computed using the divergence-free nature of the velocity {\boldmath $v$} and vorticity {\boldmath $\omega$} fields using the expansions given in Eqs.~\ref{eq.vel_model}-\ref{eq.vort_model}. We start with the marginal modes determined by the linear theory~\cite{chandrasekhar_book} and add higher-order modes generated due to the nonlinear interaction of the linear and nonlinear modes.  We select eight large-scale modes for the vertical velocity $v_{3}$: $\tilde{W}_{101}$, $\tilde{W}_{011}$, $\tilde{W}_{112}$, $\tilde{W}_{\bar{1}\bar{1}2}$, 
$\tilde{W}_{211}$, $\tilde{W}_{\bar{2}\bar{1}1}$, $\tilde{W}_{121}$, $\tilde{W}_{\bar{1}\bar{2}1}$, and twelve real modes for the vertical vorticity ${\omega}_3$: $\tilde{Z}_{101}$, $\tilde{Z}_{011}$, $\tilde{Z}_{112}$, $\tilde{Z}_{\bar{1}\bar{1}2}$, $\tilde{Z}_{211}$, $\tilde{Z}_{\bar{2}\bar{1}1}$, $\tilde{Z}_{121}$, $\tilde{Z}_{\bar{1}\bar{2}1}$, $\tilde{Z}_{\bar{1}\bar{1}0}$, $\tilde{Z}_{200}$, $\tilde{Z}_{020}$, and $\tilde{Z}_{\bar{2}\bar{2}0}$.  Notice that the vertical vorticity has two types of modes: $\tilde{Z}_{lm0}$ and $\tilde{Z}_{lmn}$. The first set of modes do not depend on the $z$ coordinates, while the second set of modes depend on the $z$ coordinate. Any conclusions without considering both types of modes may not represent nonlinear behaviour correctly.   Those vertical vorticity modes which couple linearly with the vertical velocity modes would play significant roles at higher rotation rates, and they should also be considered in a model suitable at higher values of $Ta$. Projecting the hydrodynamic system (Eqs~\ref{vorticity_model}-\ref{velocity_model}) on these modes, we get a dynamical system consisting of twenty modes. This is done using the software `MAPLE'. This model, called as model I, is then integrated using standard $RK4$ integration scheme. The model shows good qualitative agreement with DNS for low values of the Taylor number ($2 \le Ta \le 15$). For lower values of $Ta$, the model captures all the instabilities observed in DNS over a wide range of reduced Rayleigh number $r$ ($1 < r \le 1.25$) except for the chaotic behavior observed in the DNS. The comparison of the results computed from the model with those obtained from DNS for $Ta = 10$ are listed in table~\ref{table2}. 

\begin{figure}[h]
\centerline{\includegraphics[height=8 cm,width=12 cm]{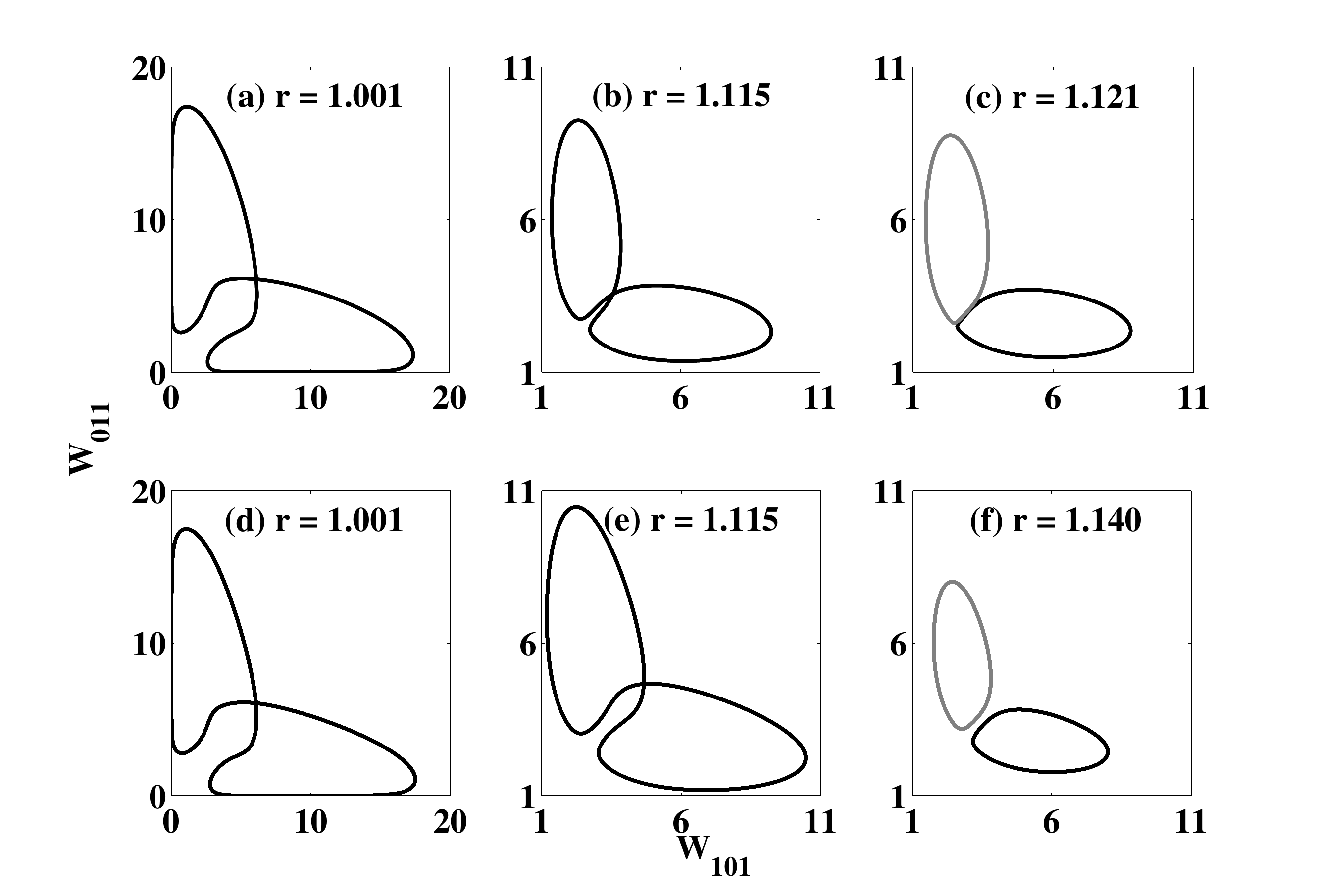}}
\caption[short]{Phase portraits for $Ta =10$ and the aspect ratio $\eta = 1$ for different values of $r$. The upper row displays the results computed from DNS for (a) $r = 1.001$, (b) $r = 1.115$, and (c) $r = 1.121$. The lower row shows the 
results obtained from the model for (d) $r = 1.001$, (e) $r = 1.115$, and (f) $r = 1.140$. The results obtained from the model are in good agreement with those computed from DNS.}
\label{phase_portrait}
\end{figure}

\subsection{The model with $L = \lambda_c$}
We now compare the results of the model for a square horizontal cross-section of each side $L = \lambda_c = 2\pi/k_c (Ta)$ with those obtained from  DNS for $Ta =10$. Figure~\ref{phase_portrait} displays the periodic dynamics of the system near the instability onset for $Ta =10$ in the $W_{101} - W_{011}$  plane.  The limit cycles in the upper row [\ref{phase_portrait}(a)-(c)], computed from DNS, describe periodic dynamics for different values of $r$. The limit cycles in the lower row [\ref{phase_portrait}(d)-(f)] are obtained from the model. The limit cycle in figure~\ref{phase_portrait}(a) corresponds to PB patterns shown in figure~\ref{sq_box} just above the instability onset ($r = 1.001$). Parts of the limit cycle lie on the $W_{101}$ and  $W_{011}$ axes. We observe straight rolls parallel to the $x$ ($y$) axis for the period $W_{011} = 0$ ($W_{101} = 0$). The limit cycle computed from the model for the same parameter value is shown in figure~\ref{phase_portrait}(d). They are in very good agreement. The variation of the  period $\tau$ of the PB patterns near the instability onset with $Ta$, obtained from the model, is shown by `$+$' mark in figure~\ref{period_limit_cycles}. The period $\tau$ computed from DNS agrees well with that obtained from the model.
 
\begin{figure}[h]
\centerline{\includegraphics[height=10 cm,width=14cm]{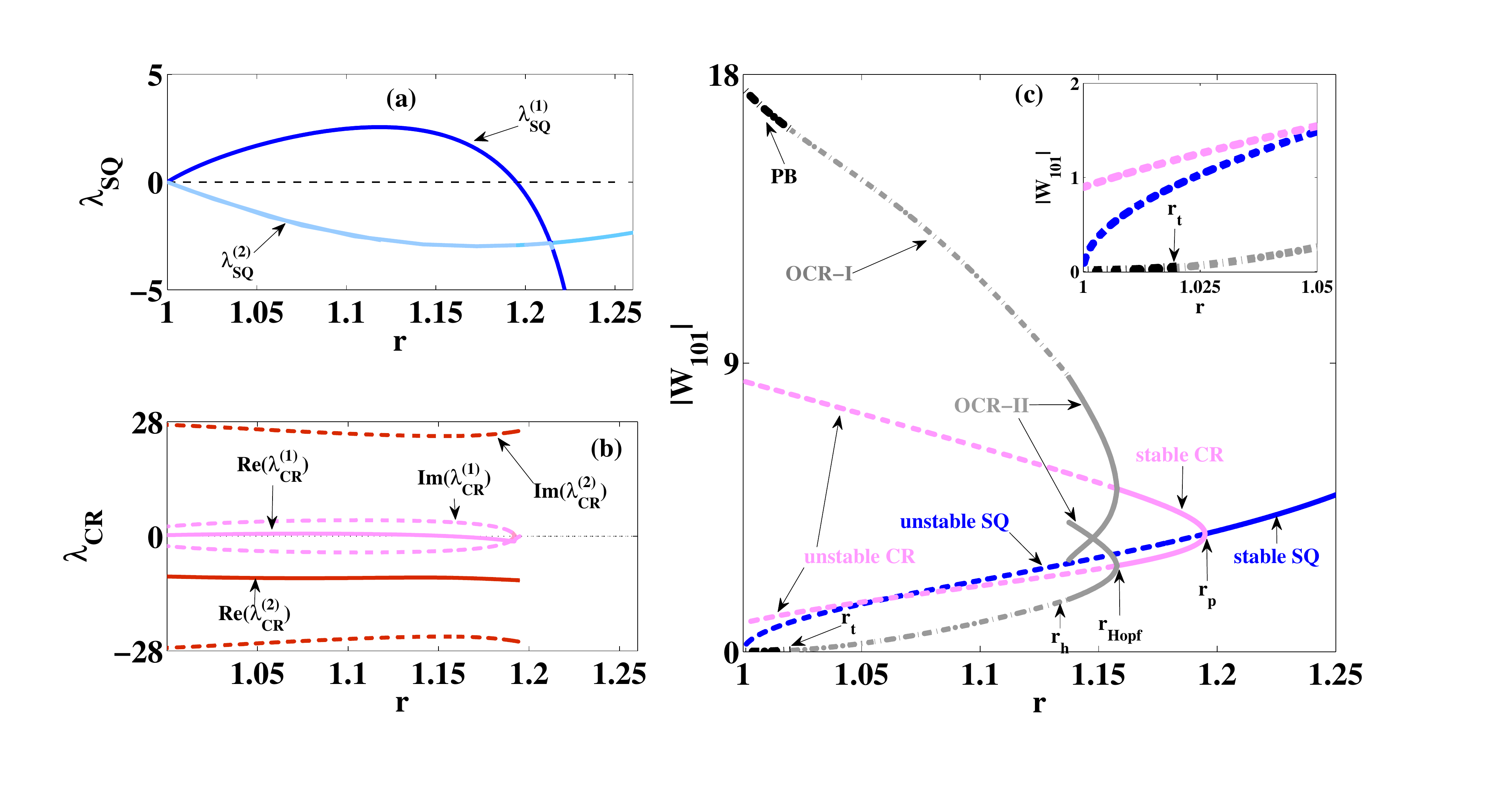}}
\caption[short]{Linear stability analysis of the nonlinear fixed points of the model for $Ta = 10$ and $\eta = 1$: (a) the variation of the two largest eigenvalues $\lambda_{SQ}^{(1)}$ (blue curve) and $\lambda_{SQ}^{(2)}$ (light blue curve) for the stationary squares (SQ) as a function of $r$. Both $\lambda_{SQ}^{(1)}$ and $\lambda_{SQ}^{(2)}$ are real. (b) the variation of the two largest eigenvalues $\lambda_{CR}^{(1)}$ (pink curves) and $\lambda_{CR}^{(2)}$ (brown curves) of cross-rolls with $r$.  Both $\lambda_{CR}^{(1)}$ and $\lambda_{CR}^{(2)}$ are complex. Their real parts are shown by solid curves and imaginary parts by broken curves. (c) Bifurcation diagram as obtained from the model. It predicts the PB, OCR-I, OCR-II, CR and SQ patterns as  $r$ is raised in small steps.}
\label{bifurcations_r}
\end{figure}

As $r$ is raised in small steps, keeping $Ta$ fixed, we observe secondary bifurcations. Figure~\ref{phase_portrait}(b) shows a limit cycle, obtained from DNS, for $r = 1.115$. Here $|W_{101}|_{max}$ $=$  $|W_{011}|_{max}$ and $|W_{101}|_{min}$ $=$  $|W_{011}|_{min}$ $\neq 0$. They describe OCR-I patterns. The trajectory forming the limit cycle appears to intersect each other because we have shown the projection of higher ($64^3$) dimensional phase space on the $W_{101} - W_{011}$  plane. With further increase in $r$, the limit cycle corresponding to OCR-I patterns breaks into two smaller limit cycles. Figure~\ref{phase_portrait}(c) displays two smaller limit cycles at $r = 1.121$.   The smaller limit cycles with $|W_{011}|_{max} \neq |W_{101}|_{max}$ describe two possible states of OCR-II patterns. The model also shows a larger limit cycle (OCR-I patterns) for $r= 1.115$ [see, figure~\ref{phase_portrait}(e)], which breaks into one of the two possible smaller limit cycles [figure~\ref{phase_portrait}(f)] for $r = 1.14$. The results of the model for the secondary and higher order instabilities are in qualitative agreement with those obtained from DNS. 

\begin{figure}[h]
\centerline{\includegraphics[height=8 cm,width=14 cm]{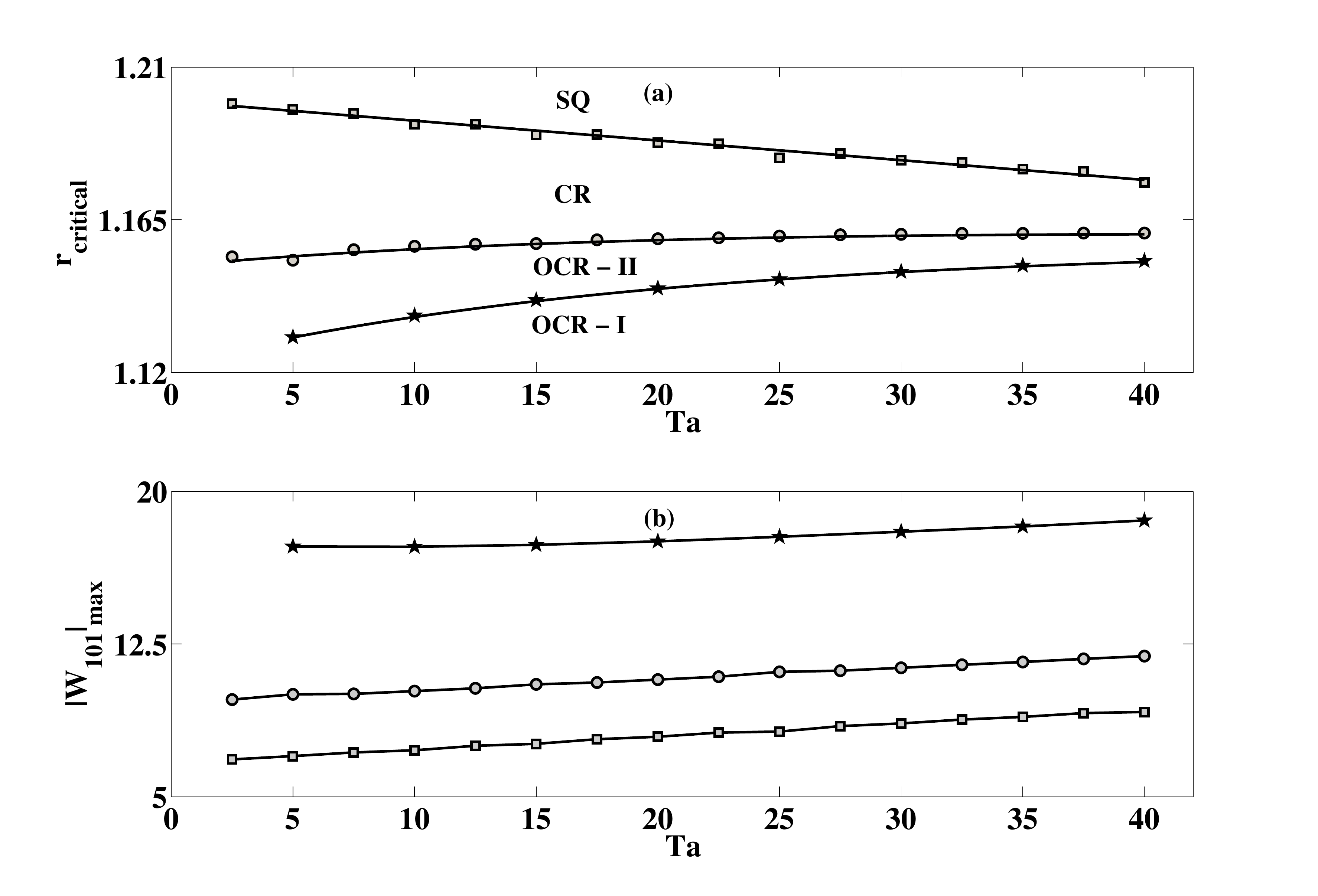}}
\caption[short]{(a) Variation of the critical value $r_{critical}$ of the reduced Rayleigh number with the Taylor number $Ta$ for different bifurcations ($\eta = 1, L = 2\pi/k_c$). The critical value of $r$ for backward homoclinic bifurcation (line marked by `$\bigstar$') and that for the backward Hopf bifurcation (marked by `O') increase slowly, while the critical value of $r$ for backward pitch-fork bifurcation (`$\Box$') decreases slowly with increasing Taylor number $Ta$. (b) Variation of the amplitude of the Fourier mode $W_{101}$ at different bifurcation points (marked by `$\bigstar$', `O' and `$\Box$') increases linearly with the Taylor number $Ta$.}
\label{rc_Ta}
\end{figure}

We now use the model to investigate the sequence of bifurcations in zero-Prandtl-number convection by varying the rotation rate. We have determined all the fixed points of the model, and carried out their linear stability analysis. The trivial fixed point (i.e., the conduction state) becomes unstable at $r = 1$ and remains unstable for all values of $r > 1$. We have found the possibility of two types of nontrivial fixed points (stationary convective states) in this model: 
the square fixed points ($|W_{101}|$ $=$ $|W_{011}|$) and the cross-roll fixed points ($|W_{101}|$ $\neq$ $|W_{011}|$). We now discuss the sequence of bifurcations obtained from the model for $Ta = 10$. The variation of the two largest eigenvalues $\lambda_{SQ}^{(1)}$ and $\lambda_{SQ}^{(2)}$ of the square fixed point with $r$ is shown in figure~\ref{bifurcations_r}(a) for $Ta = 10$. Both the eigenvalues are found to be real. They vanish at the instability onset ($r = 1$). The largest eigenvalue $\lambda_{SQ}^{(1)}$ (dark blue curve) is positive for $1 < r < 1.195$ and negative for $r \ge 1.195$. The second largest eigenvalue $\lambda_{SQ}^{(2)}$ (light blue curve) is always negative. The square fixed points are therefore stable for $r \ge 1.195$, and become saddle fixed points for $ 1 < r < 1.195$. The stationary square patterns are observed for $r \ge 1.195$ in the model for $Ta = 10$.

The cross-roll fixed points exist for $ 1 \le r \le 1.194$ for $Ta =10$. The eigenvalues of the cross-roll fixed points are found to be complex numbers.  The eigenvalues with the largest and the second largest real parts are denoted by $\lambda_{CR}^{(1)}$ and $\lambda_{CR}^{(2)}$, respectively. The variations of the real and imaginary parts of $\lambda_{CR}^{(1)}$ (pink curves) and $\lambda_{CR}^{(2)}$ (brown curves) with $r$ are shown in figure~\ref{bifurcations_r}(b). The real part of $\lambda_{CR}^{(1)}$ is found to be positive for $1 \le r < 1.158$ and negative only for $1.158 \le r \le 1.194$.  The real part of $\lambda_{CR}^{(2)}$ is always negative. The imaginary parts of the eigenvalues $\lambda_{CR}^{(1)}$ (pink dashed curves) and $\lambda_{CR}^{(2)}$ (brown dashed curves) remain finite for $r < 1.157$.  The stationary cross-roll patterns are observed in the model in the range ($1.158 \le r \le 1.194$) for $Ta = 10$. There are always two sets of cross-rolls: one with $|W_{101}|_{max}$ $>$ $|W_{011}|_{max}$ and another with $|W_{101}|_{max}$ $<$ $|W_{011}|_{max}$. They are connected by a four-fold rotational symmetry in the horizontal plane. The transition from stationary square patterns to stationary cross-rolls occurs at $r = 1.195$ through a pitchfork bifurcation. The stationary cross-rolls patterns are stable in the range $1.158 \le r \le 1.194$. They become unstable at $r = 1.157$ through the Hopf-bifurcation. The unstable cross-roll fixed points are saddle-focus for $1 \le r \le 1.157$. 

\begin{figure}[h]
 \begin{center}
\includegraphics[height=10 cm, width=14 cm]{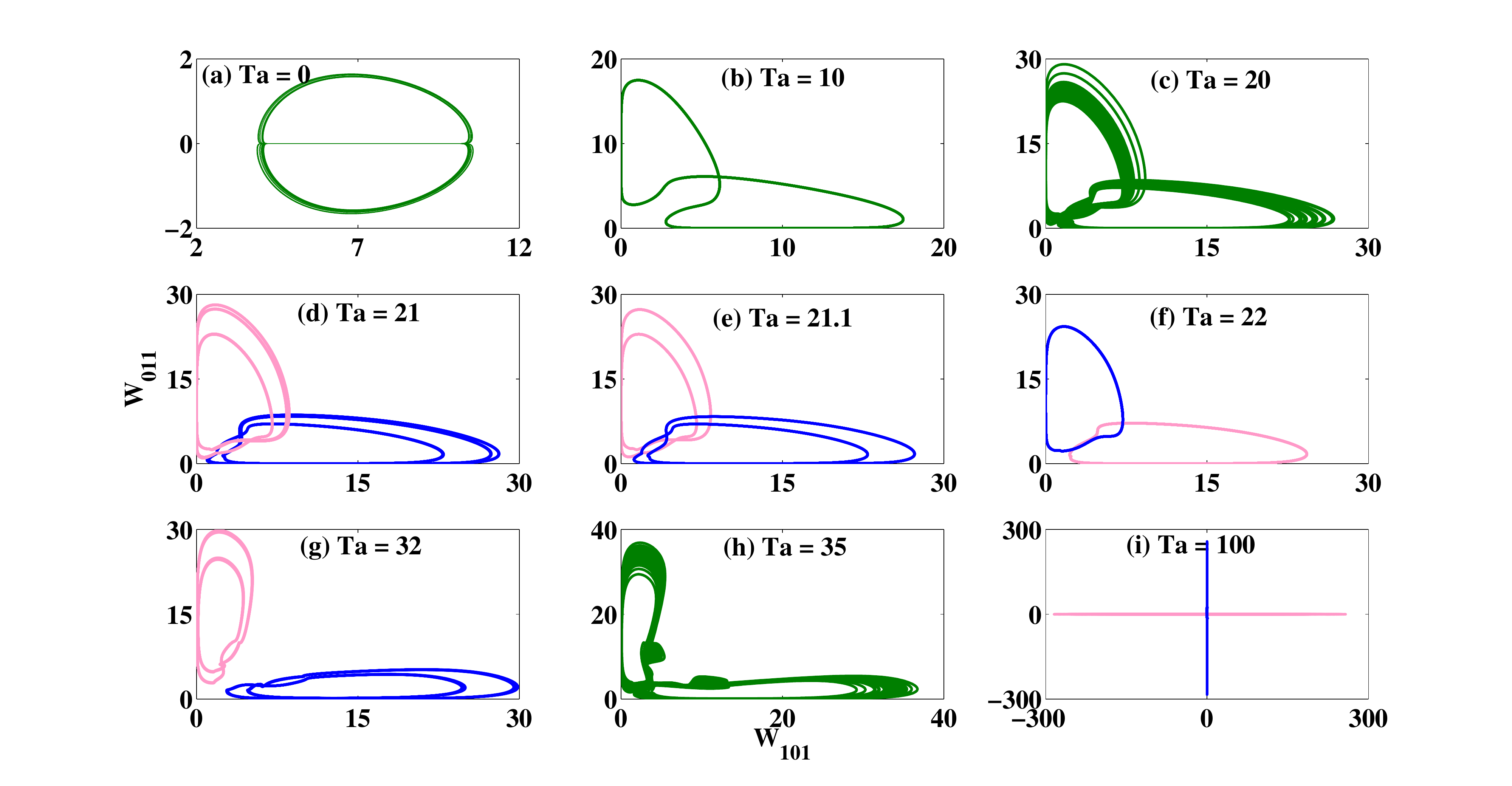}
\caption[short]{The projection of the phase portraits on the  $W_{101}-W_{011}$ plane near the primary instability ($r= 1.001$) for different values of $Ta$ ($\eta = 1, L = 2\pi/k_c$). Two sets of possible phase plots (blue and pink curves) are possible in each quadrant of the $W_{101}-W_{011}$ plane. The green curves show the glued solutions. The temporal variation of these modes are given in figure~\ref{sq_modes}.} \label{sq_phase_plots_model}
 \end{center}
\end{figure} 

\begin{figure}[h]
\centerline{\includegraphics[height=10 cm,width=10cm]{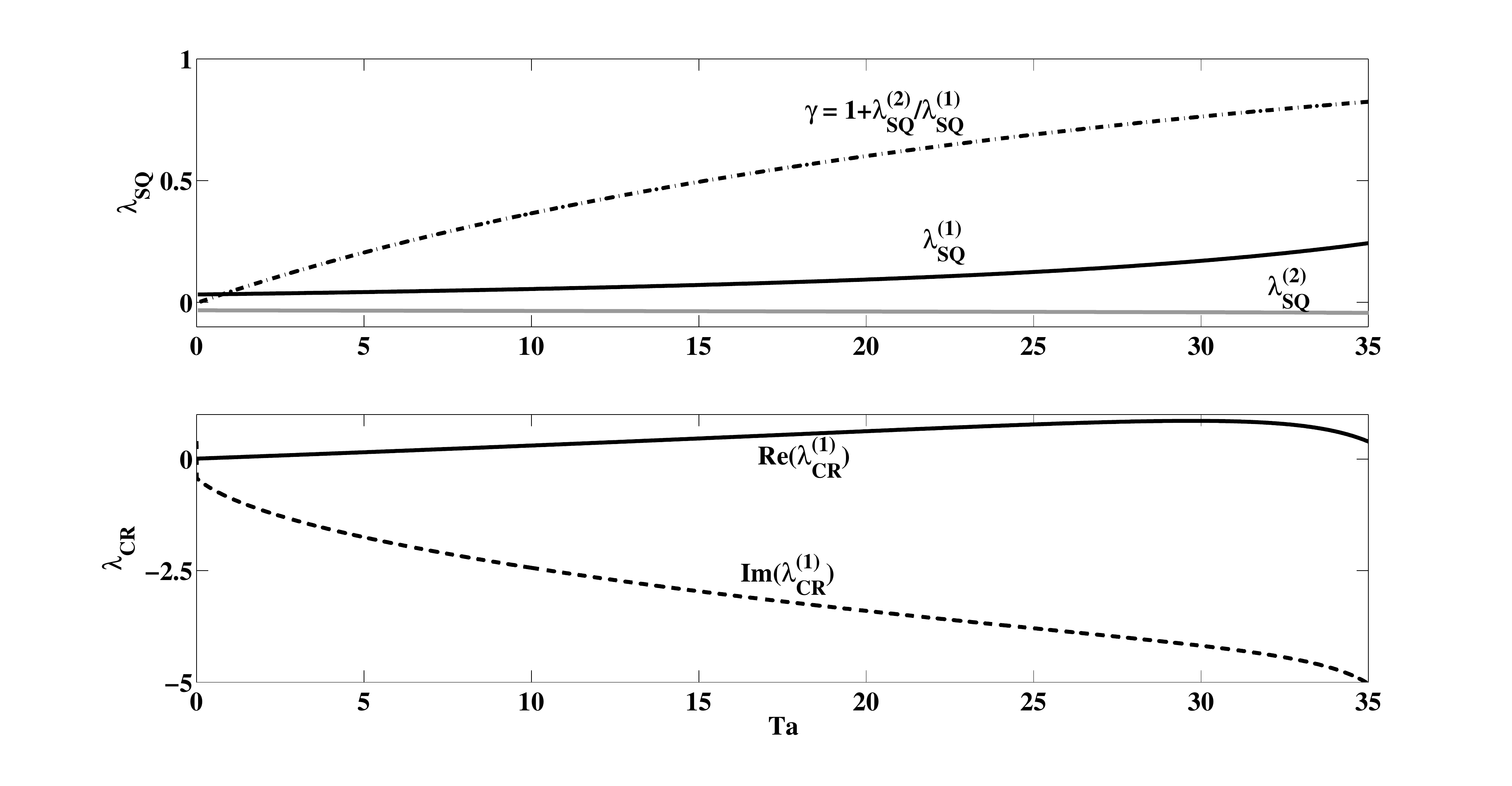}}
\caption[short]{Linear stability analysis of the nonlinear fixed points of the model near the onset ($r=1.001$) as a function of $Ta$ for $\eta = 1,  L = 2\pi/k_c$. The upper row shows the variation of the two largest eigenvalues $\lambda_{SQ}^{(1)}$ (black curve) and $\lambda_{SQ}^{(2)}$ (gray curve) for the stationary squares (SQ) as a function of $Ta$. Both $\lambda_{SQ}^{(1)}$ and $\lambda_{SQ}^{(2)}$ are real. The saddle index $\gamma = 1 + \lambda_{SQ}^{(2)} /\lambda_{SQ}^{(1)}$ (broken curve) remains always positive. Eigenvalues with the largest real parts form a complex-conjugate pair in case of cross-rolls(CR) fixed points. The variations of their real and imaginary parts with $Ta$ are shown in the lower row. The real $Re(\lambda_{CR})$ and the imaginary parts  $Im(\lambda_{CR})$ are displayed by solid and dashed black curves, respectively.}
\label{eigenvalues_Ta}
\end{figure}

\begin{figure}[h]
\centerline{\includegraphics[height=10 cm,width=14cm]{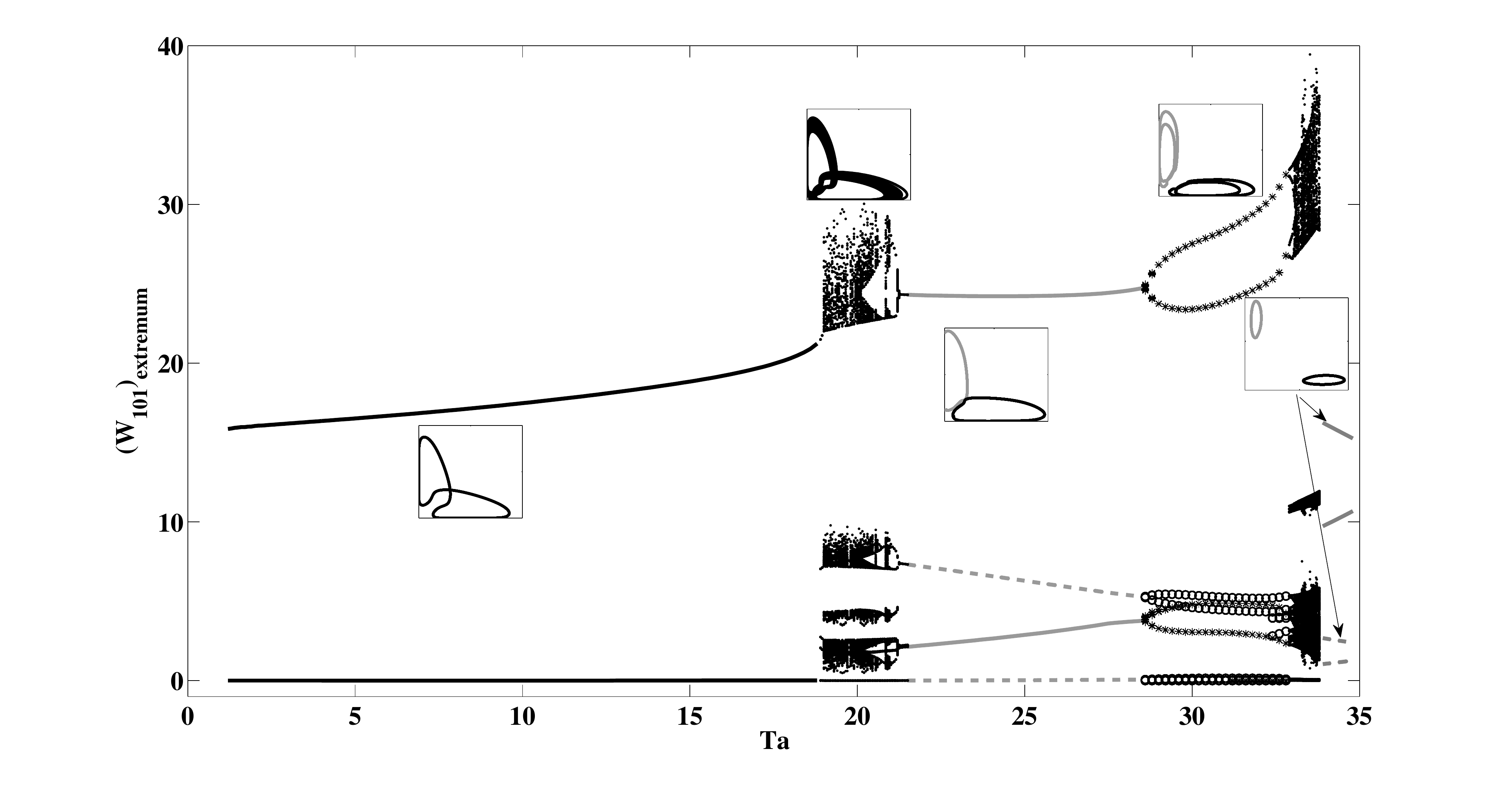}}
\caption[short]{The bifurcation diagram as a function of $Ta$, as obtained from the model-I for $r = 1.001$ and $\eta = 1$. The variations extrema of the Fourier mode $W_{101}$ with $Ta$ show glued limit cycle (black solid curves), unglued limit cycles (gray solid and broken curves), period doubling of unglued limit cycles (shown by  $\ast$ and $\circ$) and chaotic solutions (black $\bullet$). Typical phase portraits in the $W_{101}-W_{011}$ plane are also shown in square boxes for different windows in $Ta$.}
\label{bifurcations_Ta}
\end{figure}

The details of the sequence of bifurcations for $Ta =10$ are given in the bifurcation diagram [figure~\ref{bifurcations_r}(c)]. The stable and unstable stationary square patterns are denoted by blue solid and blue dashed curves, respectively. The stable and unstable stationary cross-roll patterns are shown by pink solid and pink dashed curves, respectively. No stationary solution is found to be stable for $r < 1.157$ at $Ta =10$. We have integrated the model with several different initial conditions to find out the time dependent solutions. Two sets of limit cycles are found to appear at $r = 1.157$ through Hopf bifurcation. The solid dark gray curves in the bifurcation diagram show the  maxima and the minima of both the limit cycles, which describe the two branches of oscillating cross-rolls (OCR-II) patterns. We observe a pattern with either $|W_{101}|_{max}$ $>$ $|W_{011}|_{max}$ or $|W_{101}|_{max}$ $<$ $|W_{011}|_{max}$. The size of both the limit cycles grows, as $r$ is lowered. Both the limit cycles almost touch the square (saddle) fixed points at $r = 1.137$. They simultaneously become homoclinic orbits of the same saddle fixed point. Two limit cycles are spontaneously glued together to form a larger limit cycle. The homoclinic gluing also occurs in the absence of rotation~\cite{pal_etal_2013}. The homoclinic gluing has also been observed in other systems~\cite{demeter_99, abshagen_etal_2001, peacock_2001}. The resulting  oscillating cross-rolls (OCR-I) have $|W_{101}|_{max}$ $=$ $|W_{011}|_{max}$ and $|W_{101}|_{min}$ $=$ $|W_{011}|_{min} \neq 0$. These patterns (OCR-I) exist for $1.05 < r < 1.137$. Figure~\ref{phase_portrait}(e) and \ref{phase_portrait}(f) show the single glued limit cycle and two unglued limit cycles, respectively. The glued limit cycle exists for $1 < r < 1.137$. The global maximum and minimum of the glued limit cycle are shown by dashed-dotted light gray curves in the bifurcation diagram. The minimum of the glued limit cycle vanishes as $r$ is lowered below $1.02$. In this case, the glued limit cycle describes a periodic relaxation oscillation. The amplitude of one set of rolls  remains zero for finite time, while that of the other set keeps growing. A set of new rolls gets excited, if the amplitude of the old set of rolls exceeds a critical value. This behavior continues up to the instability onset. It would be interesting to derive a normal form of amplitude equations that captures the phenomenon of critical bursting.

There is a smooth transition from oscillating cross-rolls (OCR-I) to PB of rolls in mutually perpendicular directions at $r = 1.02$. The periodic bursting of patterns is a nonlocal solution of the hydrodynamic system at the instability onset. The normal form of the amplitude equations in the $Pr \rightarrow 0$ limit with stress-free boundaries is, therefore, non-trivial. It cannot be derived using a local bifurcation analysis of the stationary conduction state.  The sequence of bifurcations for a fixed value of $Ta$ consists of a nonlocal bifurcation at the onset, an inverse homoclinic bifurcation from a glued solution to two possible nonlocal solutions, an inverse Hopf bifurcation, and an inverse pitch-fork bifurcation, as $r$ is raised in small steps. The convective patterns consist of the PB, OCR-I, OCR-II, CR and SQ patterns. The transition from the PB patterns to OCR-I patterns is smooth. Figure~\ref{rc_Ta}(a) shows the variation of the different bifurcation thresholds with $Ta$. The threshold for homoclinic bifurcation and the appearance of stationary cross-rolls increases with $Ta$, while that for the stationary square patterns decreases with increasing $Ta$. The maximum magnitude of the Fourier mode $W_{101}$ increases with $Ta$ at the onset of different bifurcations [see fig.~\ref{rc_Ta}(b)].

We have also used the model to investigate the bifurcations near onset as a function of $Ta$. The phase portraits obtained by the model for $r = 1.001$ at different values of $Ta$  are shown in figure~\ref{sq_phase_plots_model}. The model captures the qualitative behaviour observed in DNS (see figure~\ref{sq_phase_plots}). The variations of the eigenvalues of the square (SQ) and cross-rolls (CR) fixed points with $Ta$ are shown in figure~\ref{eigenvalues_Ta}. The largest eigenvalue $\lambda_{SQ}^{(1)}$ is real and positive, while the second largest eigenvalue $\lambda_{SQ}^{(2)}$ is always negative (see the upper row of figure~\ref{eigenvalues_Ta}). The square patterns are saddle fixed points near the onset ($r = 1.001$). The two largest eigenvalues of the cross-roll fixed points are found to form complex conjugate pairs for different values of $Ta$ near the onset. The variations of the real [Re($\lambda_{CR}$)] and imaginary [Im($\lambda_{CR}$)] parts  with $Ta$ are shown in the lower row of figure~\ref{eigenvalues_Ta}. The cross-rolls fixed points are unstable near the onset. The saddle index $\gamma = 1 + \lambda_{SQ}^{(2)}/\lambda_{SQ}^{(1)}$ is found to be always positive near the onset.  The nonlinear solutions of the model near onset are obtained by integrating the model. The bifurcation diagram obtained from the model with $Ta$ as a bifurcation parameter is shown in figure~\ref{bifurcations_Ta}. The solution near the primary instability can be periodic glued solution, chaotic solution, period doubling of unglued solutions, etc. The nature of the solution depends upon the value of $Ta$. The model captures the sequence of bifurcations observed in DNS quite well for smaller values of $Ta$. 

\begin{figure}[h]
\centerline{\includegraphics[height=9 cm,width=14 cm]{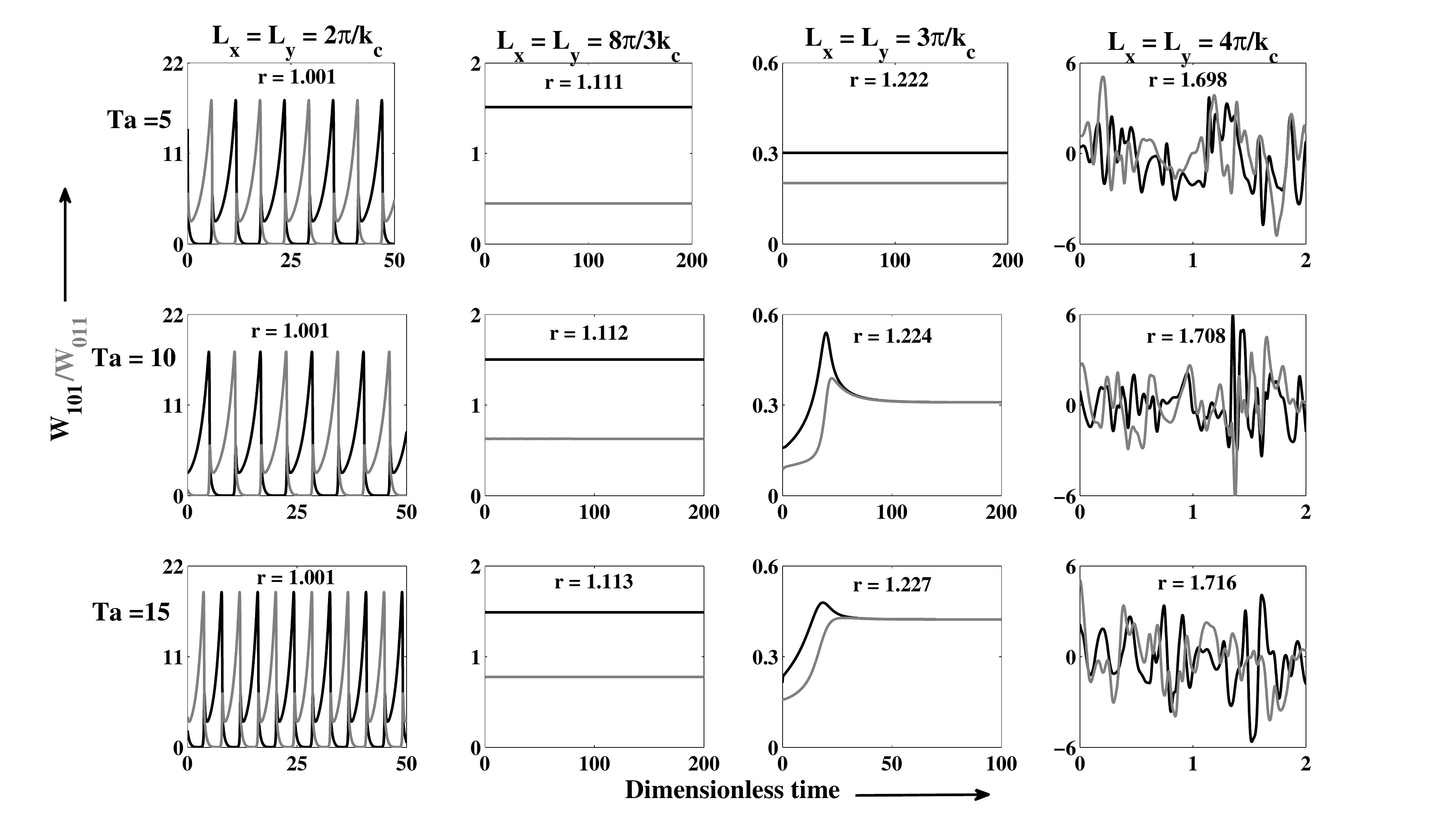}}
\caption[short]{The temporal variations of Fourier modes $W_{101}$ (black curve) and $W_{011}$ (gray curve) near the onset of convection [$Ra = 1.001 \times Ra_s (Ta, k)$] for different values of  $Ta$ in a square simulation box of side length $L$, as obtained from the model. $Ra_s (Ta, k)$ is threshold value of $Ra$ for a given value of $k < k_c$ and  $Ta$.  The temporal variations of the Fourier modes for $L = \lambda_c, 4\lambda_c/3, 3\lambda_c/2$ and $2\lambda_c$ are shown in the first, second, third and the fourth columns, respectively. Curves for $Ta = 5$, $10$, and $15$ are shown in the first, second, and the third rows, respectively.}
\label{sq_box_large}
\end{figure}

\begin{figure}[h]
\centerline{\includegraphics[height=9 cm,width=14 cm]{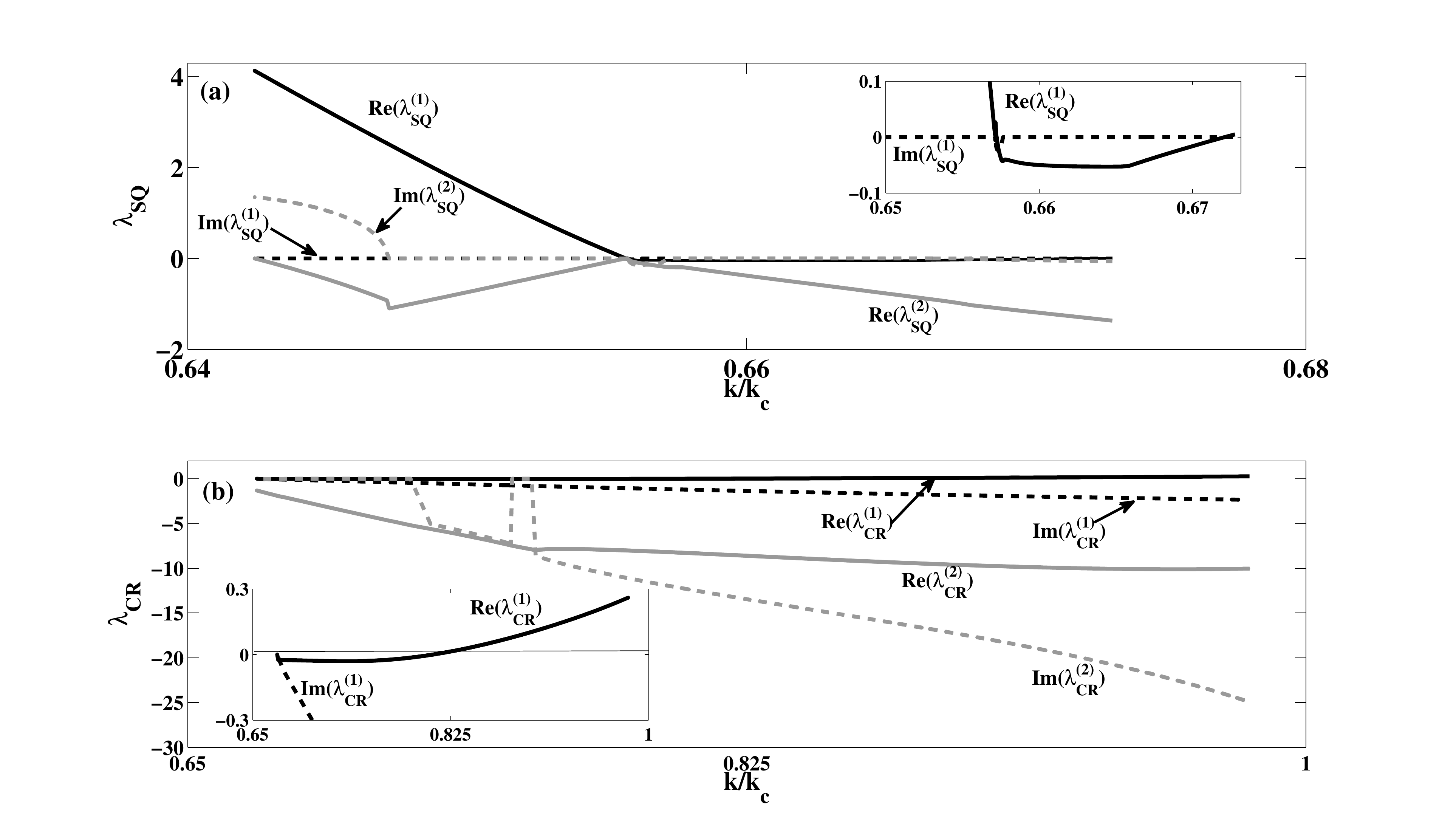}}
\caption[short]{Linear stability analysis of the nonlinear fixed points of the model near the onset ($Ra/Ra_s (k) =1.001$) as a function of $k$ for for $Ta = 10$. The upper row shows the variation of the two largest eigenvalues $\lambda_{SQ}^{(1)}$ (black curve) and $\lambda_{SQ}^{(2)}$ (gray curve) for the stationary squares (SQ) as a function of $k/k_c$. The largest eigenvalue $\lambda_{SQ}^{(1)}$ is real for the whole range of $k/k_c$. It is negative in a small range $k/k_c$ ($0.657\le k/k_c \le 0.670$).  The second largest eigenvalues $\lambda_{SQ}^{(2)}$ is complex for lower values of $k/k_c$. However, its real part is always negative. The lower row shows the variation of the two largest eigenvalues $\lambda_{CR}^{(1)}$ (black curve) and $\lambda_{CR}^{(2)}$ (gray curve) for the stationary cross-rolls as a function of $k/k_c$. Both eigenvalues are complex but the largest eigenvalue has negative real part for $0.671 \le k/k_c \le 0.810$.}
\label{eigenvalues_k}
\end{figure}

\begin{figure}[h]
\centerline{\includegraphics[height=10 cm,width=14cm]{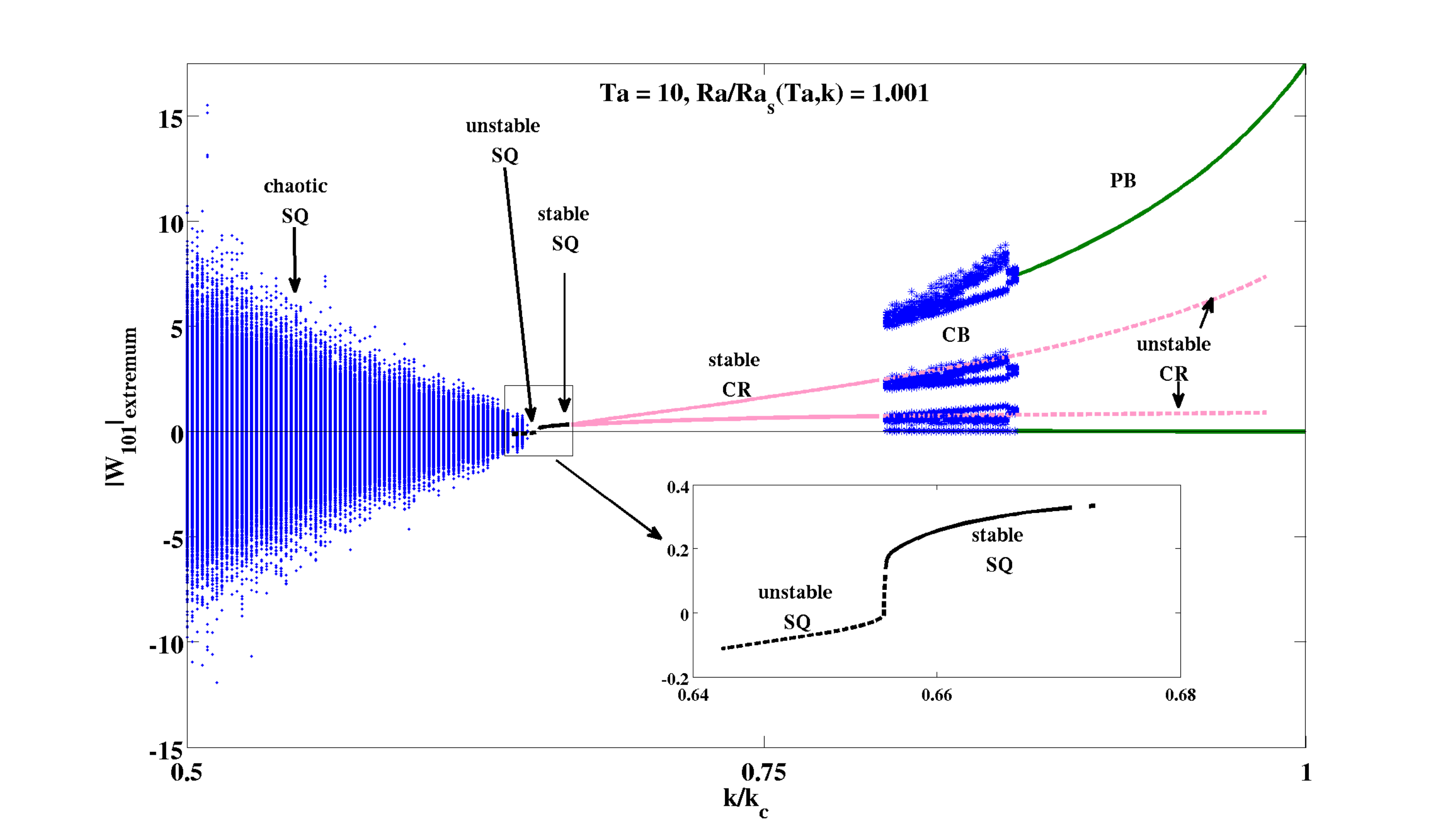}}
\caption[short]{The bifurcation diagram as a function of $k/k_c(Ta)$, as obtained from the model for $Ra/Ra_s (Ta, k) = 1.001$ and $Ta = 10$. The variation of extrema of the Fourier mode $W_{101}$ with $k/k_c$ show PB (green curves), CB (blue stars), CR (pink curves), SQ (black curves) and chaotic SQ (shown by  blue dots). The stable and unstable square fixed points are shown by solid and dashed curves, respectively (please see the inset).}
\label{bifurcations_k}
\end{figure}

\subsection{The model with $L > \lambda_c$}
The direct numerical simulations have shown that different nonlinear modes were excited in a rectangular simulation box as the ratio $\eta$ was varied. The model would require different additional modes at different rotation rates for larger simulation boxes to capture accurately even the primary instability. We have used the model  to investigate the effects in larger simulation boxes with $\eta = 1$. We have varied each side $L$ of a square cross-section of the box from $\lambda_c = 2\pi/k_c$ to $2\lambda_c$ for a given value of $Ta$.  The larger simulation boxes describe the convection away from the critical point, when the convection may grow with wave numbers $k < k_c$. The results obtained from the model near onset of convection [$Ra/Ra_s (Ta, k) = 1.001$] for different values of $Ta$ are shown in figure~\ref{sq_box_large}, where $Ra_s (Ta, k)$ is threshold value of $Ra$ for stationary convection at a given value of $k < k_c (Ta)$ and  $Ta$. Note that the convection in such cases appear at relatively higher values of $r$, although $Ra/Ra_s (Ta, k) = 1.001$ for the all cases shown in figure~\ref{sq_box_large}. We observe critical bursting of fluid patterns for $L = 2\pi/k_c = \lambda_c $, stationary cross-rolls for $L = 4\lambda_c/3$, stationary square patterns for $L = 3\lambda_c/2$, and chaotic patterns for $L = 2\lambda_c$ at the primary convection.  Stationary square patterns were observed in experiments in a container with circular cross-section by Bajaj et al.~\cite{bajaj_etal_1998}. The leading Fourier modes of the chaotic patterns for $L = 2 \lambda_c$ fluctuates about zero mean, but patterns always appear as squares. It is interesting to note that S\'{a}nchez-\'{A}lvarez et al~\cite{sanchez_etal_2005} also observed chaotically varying square patterns in DNS with realistic boundary conditions in a cylindrical container. The results of the model are in qualitative agreement with those observed in DNS. This signifies the appropriate selection of modes. The model would also be useful in investigating the effects  of variation of the horizontal cross-section of a three dimensional simulation box on pattern selection near onset. This type of model would be useful to investigate pattern dynamics near homoclinic bifurcation (e.g. in Taylor-Couette flow). It would be also applicable to investigate the interaction of stationary and oscillatory instabilities in several systems (e.g., convection in binary fluid mixtures, hydromagnetic convection, etc) near a codimension-2 point. 

We did the linear stability analysis of the square and cross-roll patterns. We computed using the model the eigenvalues of stability matrix for SQ and CR patterns as a function of $k/k_c (Ta)$ at a given value of  $Ra = 1.001 Ra_s (k, Ta)$ and $Ta$.  This investigation would reveal the effects of varying the horizontal cross-section on pattern selection. The upper row of figure~\ref{eigenvalues_k} shows the variation of the two largest eigenvalues of square fixed points with $k/k_c$ for $Ta =10$, and the lower displays the variation of the two largest eigenvalues of cross-rolls patterns with $k/k_c$. This reveals that stationary square patterns are stable in small windows of $k/k_c$ ($0.657 \le k/k_c \le 0.670$) and cross-rolls are stable for $0.671 \le k/k_c \le 0.810$. The prediction of linear stability analysis explains the stationary values of the largest Fourier modes in figure~\ref{sq_box_large}.

Figure~\ref{bifurcations_k} shows the bifurcation diagram with $k/k_c$ as bifurcation parameter for $Ra/Ra_s (k) = 1.001$ at a fixed value of $Ta = 10$. As $k/k_c$ decreases for a fixed value of $Ta$ and $Ra/ Ra_s (k)$, we observe a sequence of bifurcations. This leads to possibility of variety of patterns: PB, CB, CR, SQ and chaotic square patterns.  The model suggests that stationary square patterns are stable in a very narrow window of $k/k_c$. We could not search stationary square patterns in DNS. It is not easy to search that in parameter space of $Ta, Ra, k$. However, the model showed the possibility of selection different patterns as $L$ is varied. This may be a possible reason for observing bursting~\cite{bajaj_etal_2002} and stationary square patterns~\cite{bajaj_etal_1998} in rotating RBC.

\section{Model-II: A minimum-mode model near onset for $Ta < 10$}\label{model2}
We now try to make a minimum-mode model which captures the convective dynamics near onset of convection for smaller values of $Ta (< 10)$. The purpose is to understand the behaviour near onset as $Ta \rightarrow 0$. We expand the vertical velocity ($v_3$) and the vertical vorticity ($\omega_3$) fields as follows:
\begin{eqnarray}
v_3 &=& W_{101}\cos(kx)\sin(\pi z) + W_{011}\cos(ky)\sin(\pi z) + W_{112}\cos(kx)\cos(ky)\sin(2\pi z) \nonumber \\
&+& W_{211}\cos(2kx)\cos(ky)\sin(\pi z) + W_{121}\cos(kx)\cos(2ky)\sin(\pi z),\label{vel_8mode} \\
\omega_3 &=& Z_{110}\sin(kx)\sin(ky) + Z_{112}\sin(kx)\sin(ky)\cos(2\pi z) \nonumber \\
&+& Z_{211}\sin(2kx)\sin(ky)\cos(\pi z) + Z_{121}\sin(kx)\sin(2ky)\cos(\pi z).\label{vort_8mode}
\end{eqnarray}
\noindent where we have put $k = k_c(Ta)$, as done earlier. Projecting the hydrodynamic system (Eqs~\ref{vorticity_model}-\ref{velocity_model}) on these modes, we get a dynamical system consisting of nine modes. We then adiabatically eliminate the mode $W_{112}$, which has the highest linear decay rate. In addition, we found its amplitude of oscillation much smaller than that of other modes. This leads to the following eight-mode model:

\begin{eqnarray}
\begin{bmatrix}
\dot{X_1} \\ \dot{X_2} \end{bmatrix} &=& \epsilon(r) \begin{bmatrix}
{X_1} \\ {X_2} \end{bmatrix} - a_1 (U - 2V) \begin{bmatrix}
{X_2} \\ {-X_1} \end{bmatrix} + (a_{2} V + a_{3} U) \begin{bmatrix}
{Y_1} \\ {-Y_2} \end{bmatrix} - \dfrac{\pi}{40a_0}(U + 2V) \begin{bmatrix}
{Z_1} \\ {Z_2} \end{bmatrix}\nonumber \\ 
&-& \dfrac{c_0} {\delta(r)}\Bigg(10 \pi \begin{bmatrix}
{X_2} \\ {X_1} \end{bmatrix} - a_4 \begin{bmatrix}
{Y_1} \\ {Y_2} \end{bmatrix} + a_5 \begin{bmatrix}
{Z_1} \\ {-Z_2} \end{bmatrix}\Bigg) f({\bf{X,Y,Z}}),\label{eight_mode1}
\end{eqnarray}

\begin{eqnarray}
\begin{bmatrix}
\dot{Y_1} \\ \dot{Y_2} \end{bmatrix} &=& -\gamma(r) \begin{bmatrix}
{Y_1} \\ {Y_2} \end{bmatrix} - (b_{1}U + b_{2}V) \begin{bmatrix}
{X_1} \\ {-X_2} \end{bmatrix} - (b_{3}U + b_{4}V) \begin{bmatrix}
{Y_2} \\ {-Y_1} \end{bmatrix} + \dfrac{9\pi}{40b_0} (U + 2V) \begin{bmatrix}
{Z_2} \\ {Z_1} \end{bmatrix}\nonumber\\
&-& \dfrac{c_0}{\delta(r)} \Bigg(b_{5} \begin{bmatrix}
{X_1} \\ {X_2} \end{bmatrix} + 9\pi  \begin{bmatrix}
{Y_2} \\ {Y_1} \end{bmatrix} - 3 \begin{bmatrix}
{Z_2} \\ {-Z_1} \end{bmatrix}\Bigg) f({\bf{X,Y,Z}}),\label{eight_mode2}
\end{eqnarray}

\begin{eqnarray}
\begin{bmatrix}
\dot{Z_1} \\ \dot{Z_2} \end{bmatrix} &=& -b_0 \begin{bmatrix}
{Z_1} \\ {Z_2} \end{bmatrix} + \dfrac{\pi}{4}(5U + 4V)\begin{bmatrix}
{X_1} \\ {X_2} \end{bmatrix} - \dfrac{\pi}{40}(9U + 8V) \begin{bmatrix}
{Y_2} \\ {Y_1} \end{bmatrix} - \dfrac{9}{80}(U + 2V) \begin{bmatrix}
{Z_2} \\ {-Z_1} \end{bmatrix}\nonumber \\
&-& \dfrac{\pi c_0}{\delta(r)}\Bigg(9\begin{bmatrix}
{Z_2} \\ {Z_1} \end{bmatrix} + 10\pi \begin{bmatrix}
{X_1} \\ {-X_2} \end{bmatrix} - 27\pi \begin{bmatrix}
{Y_2} \\ {-Y_1} \end{bmatrix} \Bigg) f({\bf{X,Y,Z}}),\label{eight_mode3}\\ 
\dot{U} &=& -2c_0 U - \dfrac{\pi^2}{5}(X_{1}Y_{1} - X_{2}Y_{2}) - \dfrac{2\pi}{5} (X_{1}Z_{1} + X_{2}Z_{2}),\label{eight_mode4}\\
\dot{V} &=& -2 k_{c}^{2} V + \dfrac{\pi^2}{5}(X_{1}Y_{1} - X_{2}Y_{2}) - \dfrac{\pi}{10} (X_{1}Z_{1} + X_{2}Z_{2}) - \dfrac{\pi}{20}(Y_{1}Z_{2} + Y_{2}Z_{1}),\label{eight_mode5}\\
f({\bf{X,Y,Z}}) &=& 200\pi a_{0} X_1X_2 + 18\pi b_{0} Y_1Y_2 + 18 \pi Z_1Z_2 + a_6(X_1Z_1-X_2Z_2)\nonumber\\
&+&  a_{7}(Y_1Z_2-Y_2Z_1) + a_{8}(X_1Y_1+X_2Y_2),
\end{eqnarray}
\noindent where ${\bf{X}} = [X_1, X_2]^T \equiv [W_{101}, W_{011}]^T$, ${\bf{Y}} = [Y_1, Y_2]^T \equiv [W_{211}, W_{121}]^T$, ${\bf{Z}} = [Z_1, Z_2]^T \equiv [Z_{211}, Z_{121}]^T $, $ U \equiv Z_{112}$, and $V \equiv Z_{110}$. The coefficients  which depend on the reduced Rayleigh number are: $\epsilon (r) = [(3r - 1)k_c^{2} - \pi^{2}]$, $\gamma (r) = [b_0^{3} - 15a_0^{2} k_{c}^{2} r]/ b_0^{2}$, and $\delta(r) = 4000 (4c_0^3 - 3a_0^{2} k_{c}^2 r)$. Other coefficients which depend on $k_c(Ta)$ are defined as:
$a_0 = \pi^2+k_{c}^2$, 
$b_0 = \pi^2+5k_{c}^2$,
$c_0 = 2\pi^2 + k_{c}^2$, 
$a_1 = (\pi^{2}-k_{c}^{2})/(8a_0)$, 
$a_2 = (5k_{c}^{2}-\pi^{2})/(40a_0)$, 
$a_3 = (9\pi^{2} - 5k_{c}^{2})/(80a_0)$, 
$a_4 = \pi(11\pi^2-17k_{c}^2)/a_0$, 
$a_5 = (3\pi^2-k_{c}^2)/a_0$, $a_6 = 10(4\pi^2-k_{c}^2)$, 
$a_{7} = 3(8\pi^2-5k_{c}^2)$, $a_{8} = 20\pi(\pi^2+11k_{c}^2)$, 
$b_1 = 5(\pi^{2}-k_{c}^{2})/(8 b_0)$, 
$b_2 = (3\pi^2 + 5k_{c}^{2})/(4b_0)$, 
$b_3 = 3(13\pi^2-25k_{c}^{2})/(80b_0)$, 
$b_4 = 3(3\pi^2+25k_{c}^{2})/(40b_0)$, and
$b_5 = 10\pi (13\pi^2+5k_{c}^2)/b_0$.

\begin{figure}[h]
\centerline{\includegraphics[height=8 cm,width=16 cm]{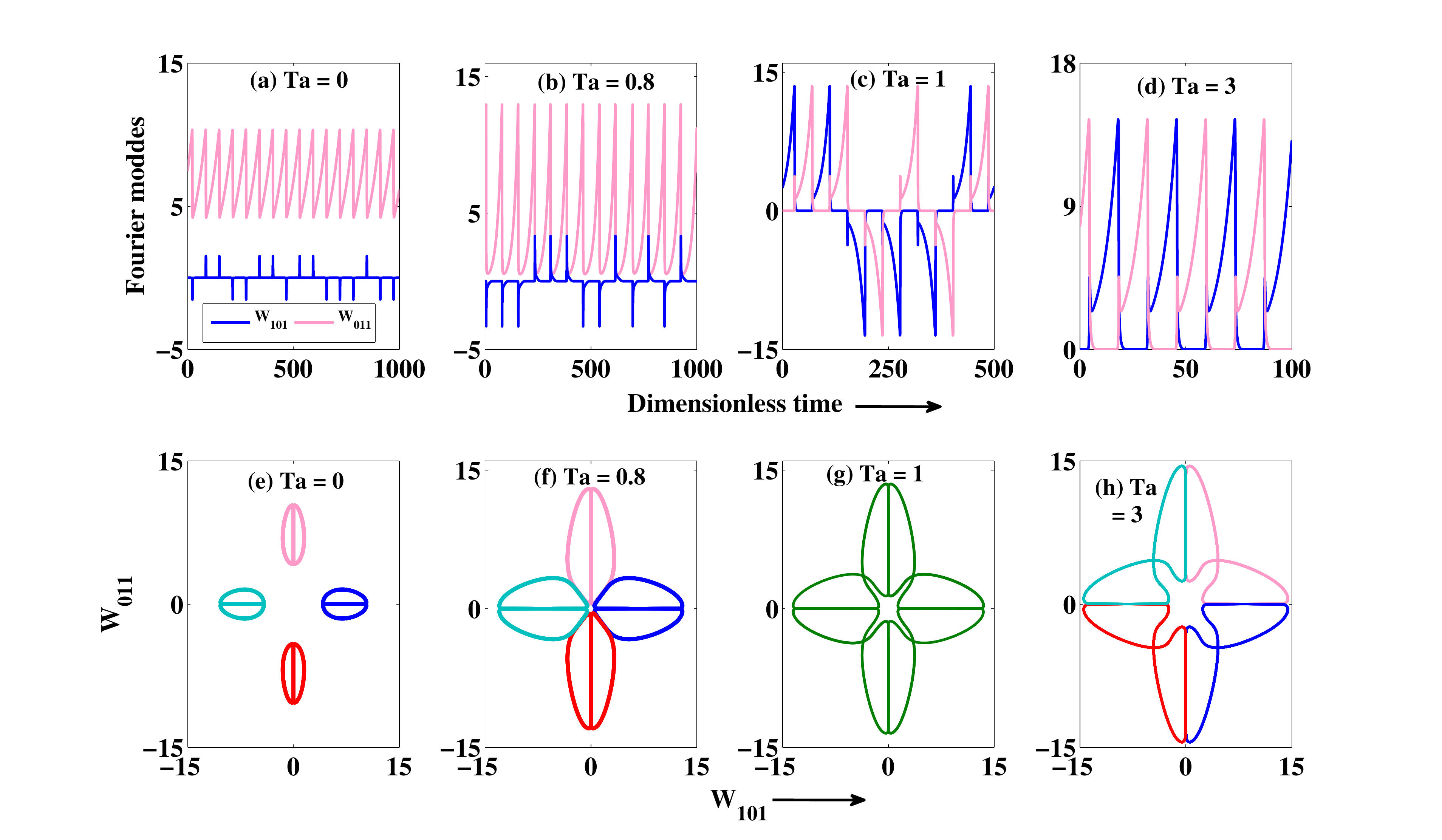}}
\caption[short]{Results obtained from the model-II near the onset $r = 1.001$ for four different values of $Ta$. First row of the figure shows the temporal variation of two largest Fourier modes $W_{101}$ and $W_{011}$ for (a) $Ta = 0$, (b) $Ta = 0.8$, (c) $Ta = 1$ and  (d) $Ta = 3$. The corresponding phase projections on $W_{101}-W_{011}$ plane are plotted in the second row of the figure.}
\label{model2_phaseplot}
\end{figure}

The elimination of any other mode leads to significantly different solutions from those observed in DNS. We therefore call this minimum-mode model as model-II.  Table~\ref{table3} shows a comparison between the results obtained from DNS and the two models (model-I and model-II) for $r = 1.001$. Though the model-II (Eqs.~\ref{eight_mode1}-\ref{eight_mode5}) shows larger deviations from the results obtained from DNS, it qualitatively captures the features observed in DNS just above onset ($r \leq 1.001$) for very small values of Taylor numbers ($0 \leq Ta \leq 10$). The temporal variations of the two largest Fourier modes $W_{101}$ and $W_{011}$ for four different values of $Ta$ obtained from model-II and the corresponding phase projections are plotted in figure~\ref{model2_phaseplot}. 

\begin{table*}[h]
\caption{Comparison of models and DNS. The global extrema of the Fourier mode $W_{101}$ and the period $\tau$ of bursting obtained from DNS, model-I and model-II are compared for different values of the Taylor number $Ta$.}
\begin{center}
  \begin{tabular}{|c||c|c|c|c|c|c|c|c|c|}
\hline
$Ta$ & \multicolumn{3}{c|}{Max($W_{101}$)} & \multicolumn{3}{c|}{Min($W_{101}$)} & \multicolumn{3}{c|}{Period of bursting ($\tau$)}\\
\cline {2-10}
  & ~DNS~ & Model-I & Model-II & ~DNS~ & Model-I & Model-II & ~DNS~ & Model-I & Model-II \\
 \hline \hline
 0	&	11.05	&	10.56	&	10.34	&	2.83		&	3.89	&	4.20	&	-	&	-  & -\\
 \hline
 3 	&	15.36	&	16.20	&	14.45	&	0	&	0	&	0	&	38.66	&	35.76		&	27.60 \\
 \hline
 5	&	15.97	&	16.52	&	14.56	&	0	&	0	&	0	&	24.35	&	22.41	&	17.42	\\
 \hline
 6	&	16.27	&	16.69	&	14.69	&	0	&	0	&	0.001	&	20.67	&	18.91	&	14.80 \\
 \hline
 8	&	16.78	&	17.06	&	14.88	&	0	&	0	&	0.005	&	15.84	&	14.45	&	11.40 \\
 \hline
 10	&	17.38	&	17.48	&	15.11	&	0.001	&	0.002	&	0.018	&	12.87	&	11.77	&	9.30 \\
 \hline
\end{tabular}
\label{table3}
 \end{center}
\end{table*}

The onset for $Ta \rightarrow 0$ limit is always found to be chaotic, while periodic solutions exist for $2 < Ta \leq 10$. These results are in agreement with the behaviour observed in DNS for $Ta \leq 10$ (please see figures~\ref{sq_modes} and ~\ref{sq_phase_plots}). The minimum-mode model, which does not capture the dynamics of the system accurately at higher values of $r$ and $Ta$, may be used to study the dynamics just near onset for smaller values of $Ta$.

\subsection{Bursting of fluid patterns near onset} 
The first row of figure~\ref{model2_phaseplot} shows the variation of the 2D roll modes $W_{101}$ and $W_{011}$ with time, and the second row gives the phase portrait in the $W_{101}-W_{011}$ plane for $r = 1.001$. The  mode $W_{011}$  varies almost periodically about a finite mean, while the mode $W_{101}$ varies chaotically with zero mean for $Ta \le 0.8$. The corresponding phase portrait is shown in the pink color in fig.~\ref{model2_phaseplot} (e).  Changing $W_{011}$ to $-W_{011}$  leads to red colored loops. Interchanging $W_{011}$ and $W_{101}$ also lead to similar dynamics. The dark blue and light blue colored loops represent them. Depending on initial conditions any one of four possible sets of chaotic solutions is possible.  Each set represents a chaotic solution here. The chaos occurs due to the random sequence of one of the most dangerous mode ($W_{101}$ or $W_{011}$). This is also an example of heteroclinic chaos with an additional feature: one of the 2D roll mode vanishes for finite time.
As $Ta$ is raised in small steps, these orbits grow in size. This behaviour persists for $0 < Ta < 0.9 $. Figure~\ref{model2_phaseplot} (b) and (f) show the temporal variation of 2D roll modes and the phase portrait respectively for $Ta = 0.8$. 

For $Ta = 0.9$, all the four chaotic solutions spontaneously merge together. Either maximum or minimum of both the 2D roll modes is zero for finite time. Figure~\ref{model2_phaseplot} (c) and (g) show such a chaotic solution and corresponding phase portrait respectively for $Ta = 1.0$. This represents a chaotic bursting~\cite{bajaj_etal_2002} of patterns at the primary instability. One set of rolls suddenly vanishes and another set of rolls perpendicular to the old sets appears. The old set of rolls is again excited  after a finite time, as the amplitude of the new set of rolls reaches a critical value.  This is different than standard K\"{u}ppers-Lortz instability, where fluid patterns do not show the phenomenon of bursting of patterns as described by Bajaj et al.~\cite{bajaj_etal_2002}. This is an example of merging of heteroclinic chaos in Rayleigh-B\'{e}nard convection with rotation. The trajectory now wanders around all four saddle fixed points. As $Ta$ is further raised, the merged chaotic solution spontaneous breaks into one of the possible four limit cycles for $Ta = 2.1$. Figure~\ref{model2_phaseplot} (d) and (h) displays the temporal variation of the two largest modes and the possibility of four possible limit cycles in different colors for $Ta =3$. Each of the 2D modes vanishes at regular interval and does grow for finite time. This leads to bursting of patterns at regular interval. This behaviour continues till $Ta =10$.

\subsection{Estimate of the time period of regular bursting of patterns} 
The mode $W_{101}$, which appear like a `delta function' in fig.~\ref{model2_phaseplot} (a) in the absence of rotation, is excited only when the mode $W_{011}$ grows close to its maximum value. It is also evident from the model-II that only the two largest Fourier modes $W_{101}$ and $W_{011}$ have positive growth rate $\epsilon$ as soon as $r > 1$. All the other modes decay linearly. Just after the mode $W_{101}$ is excited, the mode $W_{011}$ falls to its minimum value very quickly. The time of growth of rolls from its minimum intensity to the maximum is quite different from the time of decay from its maximum to minimum intensity. In the presence of even small rotation ($0 < Ta <1$), the minimum intensity of atleast one set of rolls becomes zero  [see figs.~\ref{model2_phaseplot} (b)-(d)] and remains zero for finite time. This is an example of bursting of patterns near onset. The phenomenon of bursting involves two time scales: growth time $\tau_1$ and bursting time $\tau_2$. As long as $\tau_2 \ll 0.15 \tau_1$, the bursting of fluid patterns is irregular. When $\tau_2 \sim 0.15 \tau_1$, the bursting of patterns occurs at a regular interval. The time period of the regular bursting is then estimated as: $\tau = 2\times (\tau_1 + \tau_2) \approx 2.3 \tau_1$. The factor $2$ appears because a pattern repeats after  bursting of two sets of rolls in mutually perpendicular directions [see fig.~\ref{model2_phaseplot} (d)]. We estimate the period by taking $\tau_1$ equal to the inverse of the linear growth rate. We begin with the largest linear growth rate $\sigma_{+}$ (eq.~\ref{growth_rate}). Writing $Ra = r \times Ra_c$ and setting  $k = k_c$, the expression for $\sigma_{+}$ reads as:
\begin{equation}
\sigma_{+} = \frac{-2(\pi^2 + k_c^2)^3 + k_c^2 r Ra_c + \sqrt{{(k_c^2 r Ra_c)^2 - 4 {\pi}^2 (\pi^2 + k_c^2)^3 Ta}}}{2(\pi^2 + k_c^2)^2}. \label{growth_rate1}
\end{equation}
For smaller values of Taylor numbers ($Ta < 10$), $(k_{c}^2 r Ra_{c})^2 >> 4 {\pi}^2 (\pi^2 + k_{c}^2)^3 Ta$ near the onset ($r \approx 1$). 
Ignoring $4 {\pi}^2 (\pi^2 + k_{c}^2)^3 Ta$ with respect to $(k_{c}^2 r Ra_{c})^2$ in the expression for $\sigma_{+}$ and using Eq.~\ref{modified_Rc} leads to $\sigma_{+} \rightarrow \epsilon (r) = (3r - 1)k_c^{2}(Ta) - \pi^{2}$. Binomially expanding the expression for $k_c (Ta)$ (Eq.~\ref{kc}) in $Ta$ and retaining terms upto cubic order yield
\begin{equation}
\epsilon(r,Ta) = \dfrac{3 \pi^{2}(r-1)}{2} + \dfrac{2(3r-1)}{9 \pi^{2}}Ta - \dfrac{8(3r-1)}{9 \pi^{6}}Ta^{2} - \dfrac{(3r-1)}{9 \pi^{10}}Ta^{3} + \mathcal{O}(Ta^4).
\label{growth_rate_model}
\end{equation}

\begin{figure}[h]
\centerline{\includegraphics[height=6 cm,width=16 cm]{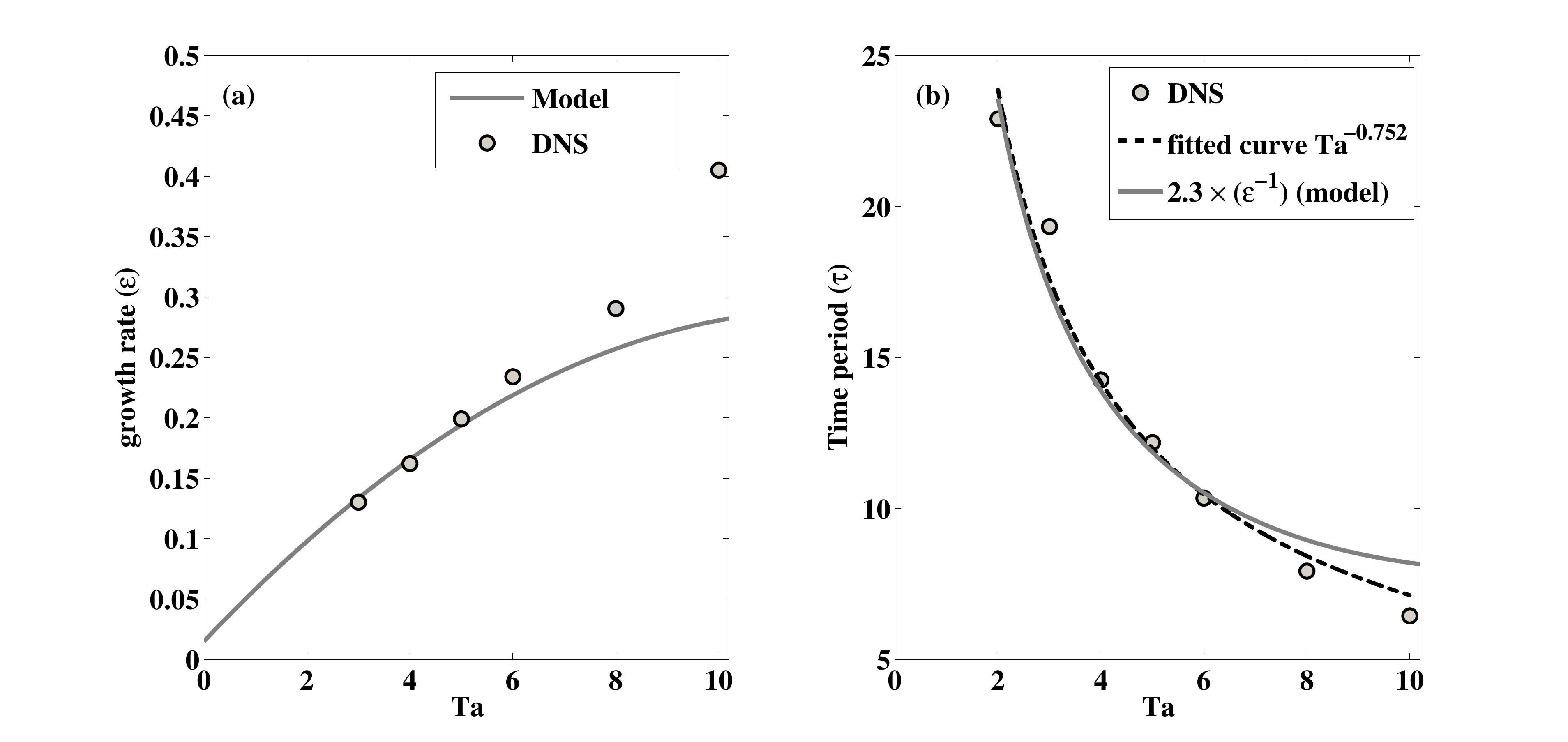}}
\caption[short]{Comparison of growth rates and time periods ($\tau$) of the two largest modes $W_{101}$ or $W_{011}$ at the onset ($r = 1.001$) as obtained from the model II using eq.~\ref{growth_rate_model} and DNS.}
\label{growth_rate_model_DNS}
\end{figure}
The time constant for the exponential growth of the rolls is $\tau_1 = 1/\epsilon(r)$, and the period of regular bursting is $\tau \approx 2.3 /\epsilon(r)$. The linear growth rate of the largest mode just above the onset is therefore given as:  $\epsilon(r \rightarrow 1,Ta)$ $=$  $4 Ta/(9\pi^2)  - 16 Ta^{2}/(9\pi^6) - 2  Ta^{3}/(9 \pi^{10})$. Figure~\ref{growth_rate_model_DNS} (a) shows a comparison of the linear growth rate $\epsilon (r)$ of the model-II (gray solid curve) and the exponential growth rate extracted from DNS (circles). The deviation of growth rates obtained from the model-II  and DNS become clearly visible for $Ta > 7$. Figure~\ref{growth_rate_model_DNS} (b) shows the comparison of the time periods of bursting of patterns for $r = 1.001$ obtained from model-II (gray solid curve) and DNS (circles). A curve (dashed line) fitted to the DNS data shows $\tau \sim Ta^{-0.752}$ for smaller values of $Ta$. Actually the period $\tau$ involves polynomial in $Ta$. The model also shows that the period of oscillation would be finite for any $Ta$, if $r-1$ is not negligible. The limit $Ta \rightarrow 0$ is not meaningful as periodic bursting is not observed for $Ta \le 2$. Figure~\ref{period_limit_cycles} showed that the period $\tau$ of bursting near onset for larger range of $Ta$ ($2 < Ta \le 40$), after omitting the data points for chaotic bursts,  scaled with $Ta$ as $Ta^{-0.784}$. The vertical vorticity modes linearly coupled to the vertical velocity also become significant for relatively higher values of $Ta$, and the linear growth rate is modified. This in turn changes the time constant $\tau_1$.  The frequency of oscillatory convection (Eq.~\ref{W0}) is imaginary for $Ta < 328$. The bursting of patterns is therefore not the consequence of the oscillatory convection at onset. It is due the instability of a set of growing rolls due to the generation of vertical vorticity, as the amplitude of the 2D rolls crosses a critical value.  

\subsection{Bifurcation of bursting patterns} 
\begin{figure}[h]
\centerline{\includegraphics[height=10 cm,width=14 cm]{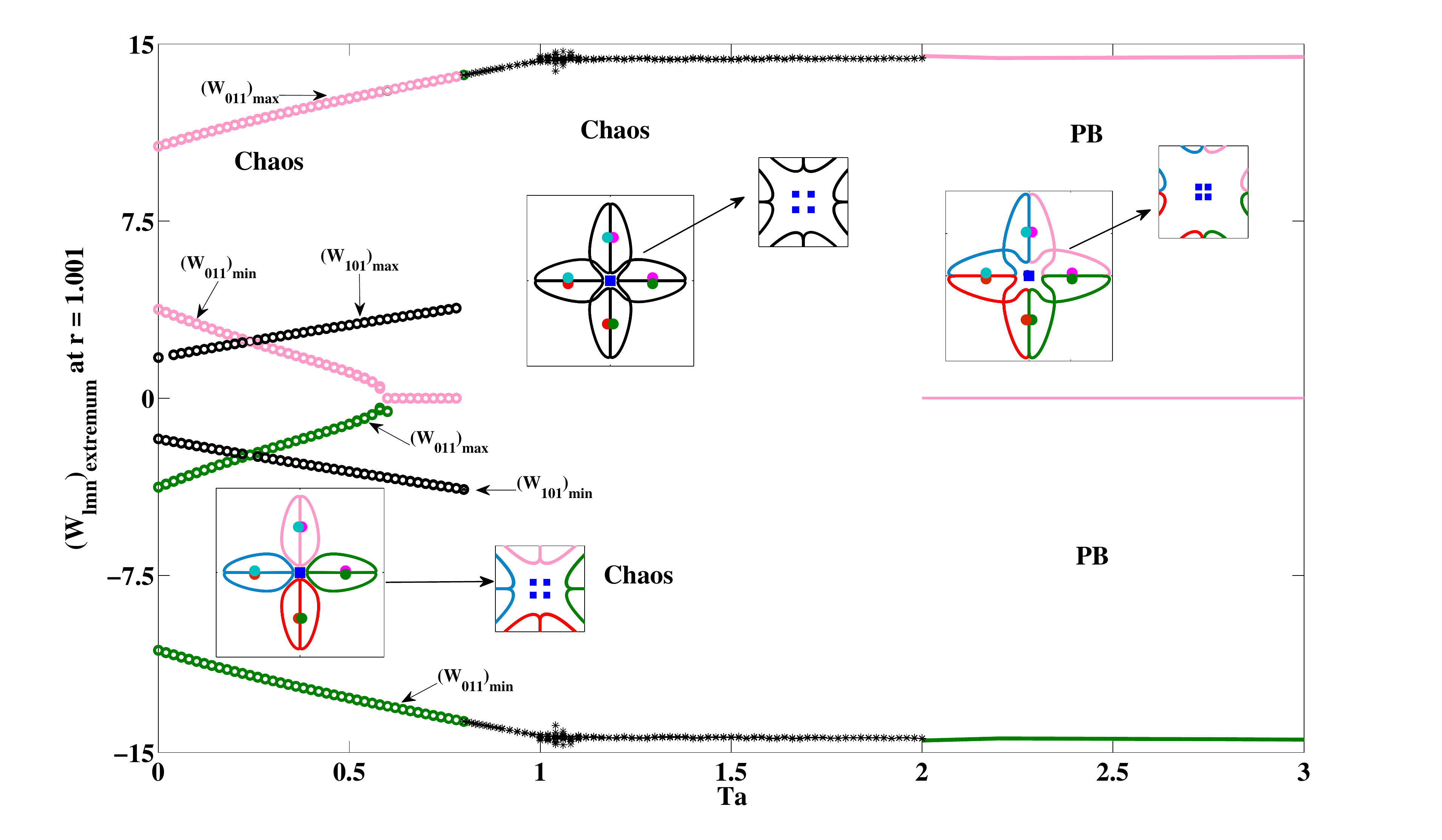}}
\caption[short]{Bifurcation diagram at the onset ($r = 1.001$) obtained from the minimum-mode model (model-II) with $Ta$ as the bifurcation parameter. Two possible sets of extrema for the mode $W_{011}$ are marked as pink and green  `$\circ$' for $0 \leq Ta \leq 0.8$.  The extrema for mode $W_{101}$ for the same range of $Ta$ are shown in black `$\circ$'. For $0.8 < Ta \leq 2$, both the modes have only one set of identical extrema marked by black  `$\ast$'.  For $2 \le Ta < 15$, two sets of identical extrema (pink and green curves) are possible for both the modes, and one of them is always zero.  The typical phase portraits for different solutions are shown in insets. Blue squares represent saddle squares and dots in different colors represent unstable (saddle focus) cross-rolls.}
\label{bifurcation_model2}
\end{figure}

Figure~\ref{bifurcation_model2} displays the bifurcation of solutions obtained from model-II near the onset of convection for $0 \le Ta \le 3$, as $Ta$ is varied in small steps. Two sets of possible extrema for the mode $W_{011}$, marked by pink and green circles, exist for $0 \le Ta < 0.6$, while only one set of extrema exists for the mode $W_{101}$ in this range of $Ta$. Typical phase portraits in the $W_{101}-W_{011}$ plane are shown in the first inset by pink and green loops for $0 \le Ta < 0.6$. Interchange of modes $W_{011}$ and $W_{101}$  give another pair of solutions. The corresponding phase portraits are shown in blue and red loops in the inset. Blue  squares in the insets represent four saddle square fixed points, one in each quadrant of the $W_{101}-W_{011}$ plane. Dots in different color represent eight saddle focus (cross-rolls) fixed points, two in each quadrant of the $W_{101}-W_{011}$ plane. Each loop represents a possible chaotic solution, as $W_{101}$ varies chaotically with time. Each loop comes close to each other as $Ta$ is raised slowly. For $0.6 < Ta < 0.8$, the minima of of the upper branch $W_{011}$ (shown by pink circles) and the maxima of the lower branch (shown by green circles) become zero. Each loop touches the a pair of saddle fixed points.  We have four distinct sets of possible heteroclinic chaotic solutions for $0.6 < Ta \le 0.8$. This is when the fluid patterns show the phenomena of bursting. As the value of $Ta$ is raised above $0.8$, four distinct sets of chaotic solutions merge together to form another chaotic solution. The typical phase portrait of this situation is shown by black curves in the second inset.  The modes $W_{011}$ and $W_{101}$ have identical extrema but they have a phase difference [see Fig.~\ref{model2_phaseplot} (c)]. As the value of $Ta$ is raised above $2$, the merged chaotic solution spontaneous break into one of four possible limit cycles (periodic solution), each shown in a different color. This happens when the loops made by chaotic trajectories just touch the square saddle fixed points. This represents an inverse heteroclinic bifurcation. After this bifurcation, each limit cycle is confined to only one quadrant of the $W_{101}-W_{011}$ plane. There is a transition from irregular bursting of patterns to regular bursting of patterns.  Bajaj et al.~\cite{bajaj_etal_2002} had also reported the periodic bursting of fluid patterns in their experiments in low-Prandtl-number fluids with rotation. Each limit cycle is closer to only one of of the four saddle square fixed points. This minimum-mode model is not good enough to capture dynamical behaviour away from the onset. Bifurcation diagram shown in Fig.~\ref{r_Ta_space} based on model-I shows that the increase in  $r$ at a fixed value of $Ta$ in certain range leads to homoclinic merging (OCR-I solution). 

The projection of the phase space in the $W_{101}-W_{101}$ plane  (Fig.~\ref{bifurcation_model2}) explains the exponential growth of one of 2D roll modes ($W_{101}$ and $W_{101}$) to a large value followed by sudden fall to zero value. For $r = 1$ ($\epsilon (r=1) = 0$), all the points on the $W_{101}$ and $W_{101}$ axes are fixed points for smaller values of $Ta$ ($Ta \rightarrow 0$). The origin represents the conduction state, while other fixed points represent 2D rolls. As $r$ is raised slightly above unity, all these fixed points become unstable. In addition, four saddle fixed points describing unstable square patterns (blue squares) and eight saddle foci describing unstable cross-rolls  patterns (dots in different colors) are generated. The four saddle fixed points surround the origin, while a pair of saddle foci are located very closely on two sides of any axis in positive as well as negative directions. The stable and unstable manifolds of these fixed points now decide the dynamics of any trajectory in the phase space. For example, a pair of saddle foci on the two sides of the positive $W_{011}$ axis guide any trajectory starting near origin on this axis to keep moving on the axis in the positive direction to a large positive value. As soon as it crosses a critical values, the vertical vorticity is excited through nonlinearity. This transfers the energy of energy from 2D roll mode $W_{011}$ to another 2D mode $W_{101}$ in a very short time. The trajectory then moves away from the $W_{011}$ axis, makes a loop around the unstable cross-roll fixed point and returns near the origin on the $W_{101}$ axis. The rolls patterns parallel to the $y$ axis disappears and a new set of rolls parallel to the $x$ axis stars growing.  The trajectories form, depending upon the value of $Ta$, either an open but bounded or a closed heteroclinic orbit. They correspond to irregular (chaotic) or regular (periodic) bursting of patterns. As observed in DNS and  model-I, the orbit may become either an open or a closed homoclinic orbit for $Ta \ge 10$ (see Fig.~\ref{phase_portrait}) near onset. 

\section{Conclusions}
We have investigated the convective patterns at primary instability using direct numerical simulations in zero-Prandtl-number Rayleigh-B\'{e}nard convection with uniform rotation. The direct numerical simulations have been done in the viscous regime with stress-free boundary conditions. The convective patterns at the instability onset are found to be time-dependent in this limit. The phenomenon of bursting of fluid patterns has been observed at the onset for small values of $1 < Ta < 40$ in a simulation box with square horizontal cross-section ($\eta = 1$) with each side $L= \lambda_c$. Bursts of large-amplitude convection appeared and then disappeared in mutually perpendicular directions. The bursting of patterns occurs at irregular intervals for certain windows of $Ta$ ($2 \le Ta \le 10$ and $28 \le Ta \le 40$), while it occurs at a regular interval in some other windows of $Ta$ ($0 \le Ta \le 10$ and $10 < Ta < 28$). Increasing $r$ at a fixed value of $Ta$ ($5 \le Ta \le 40$) leads to a series of interesting fluid patterns: merging of two  limit cycles (oscillating cross-rolls, OCR-I), spontaneous breaking of the merged limit cycle into one of the possible smaller limit cycles (OCR-II), stationary cross-rolls, and stationary squares. 

The pattern dynamics at the onset is different in rectangular simulation boxes.  Periodic  wavy rolls are observed at the onset for lower values of $Ta$, while  KL patterns are observed at moderate values of $Ta$ for $1/\sqrt{3} \le \eta \le 2$. There is a transition from to GKL patterns near a bicritical point. In larger simulation boxes with rectangular horizontal cross-section ($\eta \ge 4$), KL pattern are observed near onset even at smaller values of $Ta$. The K\"{u}ppers-Lortz patterns (KL and GKL) also show irregular bursting occasionally. Temporally quasiperiodic patterns are observed in the oscillatory regime ($Ta > 550$) for $\eta = 1$ and $2$. In the oscillatory regime ($Ta > 550$) near a bicritical point ($Ra_c (Ta) = Ra_{\circ}$), temporally quasiperiodic or chaotic cross-rolls ($|W_{101}| \neq |W_{011}|$) have been observed for $\eta =1$ and $2$. They do not show the phenomenon of bursting of patterns.  

We have also presented two low-dimensional models for a square simulation box ($\eta = 1$). A twenty-mode model (model-I) captures the sequence of secondary and higher-order instabilities qualitatively well for $L = 2\pi/k_c$ at smaller rotation rates ($Ta < 50$). The model clearly reveals the origin of bursting. When the amplitude of a set of rolls exceeds a large critical value, the vertical vorticity modes are nonlinearly excited. The bursting involves a nonlocal bifurcation at the primary instability for smaller values of $Ta$. This also shows a new possibility of large variation of the roll amplitude at a supercritical bifurcation. Increasing $r$ leads to a series of interesting instabilities including inverse homoclinic, inverse Hopf and inverse pitchfork bifurcations. The model also captures the sequence of bifurcations qualitatively at the primary instability as $Ta$ is varied. The model also shows the possibility interesting bifurcations as the parameter $k/k_c$ is varied keeping $Ta$, $r$ and $\eta$ fixed. It shows the possibility of stationary cross-rolls for $L= 8\pi/3k_c$ and stationary square patterns for $L = 3\pi/k_c$. This suggests that selection of convective patterns near the onset in rotating convection may be significantly influenced by the size of the horizontal cross-section of the container. The model shows stationary patterns instead of K\"{u}ppers-Lortz patterns at the onset by varying the horizontal cross-section slightly with fixed values of all other parameters. This type of model would be useful to understand the unfolding of bifurcations near a bicritical point. A minimum-mode model (model-II), effective just above the onset and for $Ta < 10$ shows interesting dynamics. The model shows heteroclinic chaotic solutions, merging of heteroclinic chaotic solutions, and spontaneous breaking of the merged chaotic solution into a limit cycle solution at the primary instability for $Ta < 10$.

\section*{Acknowledgment}
We benefited from fruitful discussions with Pinaki Pal, Hirdesh K. Pharasi,  Arnab Basak, and Deepesh Kumar. 

\end{document}